\documentclass[aps,superscriptaddress, nofootinbib]{revtex4}
\usepackage{amssymb,amsmath,epsfig}
\usepackage[colorlinks=true, pdfstartview=FitV, linkcolor=blue, citecolor=blue, urlcolor=magenta, breaklinks=true]{hyperref}
\usepackage{graphicx}
\usepackage{epstopdf}
\usepackage{subfig}

\usepackage{xcolor}
\usepackage{adjustbox}
\usepackage{graphicx}
\usepackage{booktabs}       % professional-quality tables
\usepackage{amsfonts}       % blackboard math symbols
\usepackage{nicefrac}       % compact symbols for 1/2, etc.
\usepackage{orcidlink}
\usepackage{cleveref}

\usepackage{xcolor}

\begin{document}

\title{Glueballs Confinement and Cosmological Phase Transitions}

\author{Adamu Issifu \orcidlink{0000-0002-2843-835X}} 
\email{ai@academico.ufpb.br}
\affiliation{Departamento de F\'isica, Instituto Tecnol\'ogico de Aeron\'autica, DCTA, 12228-900, S\~ao Jos\'e dos Campos, SP, Brazil} 
\affiliation{Laborat\'orio de Computa\c c\~ao Cient\'ifica Avan\c cada e Modelamento (Lab-CCAM),  SP, Brazil}

\author{Julio C. M. Rocha \orcidlink{0000-0003-4515-9245}}\email{julio.rocha@servidor.uepb.edu.br}
\affiliation{Departamento de F\'isica, Universidade Estadual da Para\'iba, 58233-000, Araruna, Para\'iba, Brazil}

% \author{Stephen Owusu}\email{stevenowusu@pos.if.ufrj.br}
% \affiliation{Instituto de F\'isica, Universidade Federal do Rio de Janeiro,
% Caixa Postal 68528, 21941-972  Rio de Janeiro, Brazil}

\author{Francisco A. Brito \orcidlink{0000-0001-9465-6868}}\email{fabrito@df.ufcg.edu.br}
\affiliation{Departamento de F\'isica, Universidade Federal da Para\'iba, 
Caixa Postal 5008, 58051-970 Jo\~ao Pessoa, Para\'iba, Brazil}
\affiliation{Departamento de F\'{\i}sica, Universidade Federal de Campina Grande
Caixa Postal 10071, 58429-900 Campina Grande, Para\'{\i}ba, Brazil}

\author{Tobias Frederico \orcidlink{0000-0002-5497-5490}} 
\email{tobias@ita.br}

\affiliation{Departamento de F\'isica, Instituto Tecnol\'ogico de Aeron\'autica, DCTA, 12228-900, S\~ao Jos\'e dos Campos, SP, Brazil} 
\affiliation{Laborat\'orio de Computa\c c\~ao Cient\'ifica Avan\c cada e Modelamento (Lab-CCAM),  SP, Brazil}

\begin{abstract}

We develop a unified framework in which the dynamics of a scalar glueball field, originating from phenomenological nonperturbative QCD confinement, simultaneously governs the deconfinement transition of strongly interacting matter and drives cosmological inflation. Starting from a temperature-dependent effective potential $V_{eff}(\phi, T)$, we show that the glueball mass vanishes at a critical temperature $T_{c\phi}$, signaling a first-order phase transition characterized by supercooling and a transient metastable vacuum. In the high-temperature regime $T > T_{c\phi}$, the deconfined phase naturally produces an exponential expansion of the scale factor, providing the correct conditions for inflation. By computing the slow-roll parameters and the resulting spectral index $n_s$, tensor-to-scalar ratio $r_s$, and running $\alpha_s$, we confront the model with the Planck observations. The predicted values of $n_s$ and $r_s$ fall within the Planck confidence contours for a broad and physically motivated range of the parameter $\gamma$ and for $N \approx 50\text{--}60$ e-folds. A distinctive linear relation, $r_s = 4(1-n_s)-72\gamma$, emerges as a testable signature of the model. Normalization to the observed scalar amplitude further constrains the thermal correction parameter $\sigma^2$ and the coupling $\gamma$, linking cosmological data directly to QCD-scale dynamics. These results demonstrate that a confinement-inspired potential can naturally reproduce the observed inflationary phenomenology and offer a novel bridge between early-universe cosmology and the nonperturbative sector of QCD.

\end{abstract}
\maketitle
\pretolerance10000

\section{Introduction}
The concept of {\it spontaneous symmetry breaking} (SSB) is an important subject in modern-day particle physics theories \cite{Slavnov, Taylor, Okun, Polyakov, Langacker, Nilles}, particularly, after the historic discovery of the Higgs boson. The concept plays a significant role in the unification of the fundamental forces as well. It also makes it possible to differentiate between various types of fundamental interactions among particles. The classification of strong and weak interactions was accomplished using gauge theories with SSB and asymptotic freedom \cite{Kolb, Collins, Linde5}. The increasing progress in elementary particle theory has enhanced the understanding of particle interactions even at sufficiently high energies. That led to advances in the theories of superdense matter, such as the high baryon density regime of the early universe, the inner core of the neutron star, and so on. Thus, asymptotically free theories opened the path for studying matter with densities of about $80$ times greater than the nuclear density \cite{Linde2}. The interest in applying SSB theories to cosmology is motivated by the theoretical projection that at higher temperatures, spontaneously broken symmetries are restored \cite{Kirzhnits, Kirzhnits1, Linde4}. Temperature modification (high or low) leads to a corresponding phase transition, resulting in a significant modification in matter phases \cite{Linde3, Weinberg, Dolan}. Thus, the phase transitions during the evolution of the early universe can be linked to the spontaneous breakdown of gauge symmetry \cite{Mukhanov}. 

We investigate strong interactions between glueballs and their role in the evolution of the early universe using a simple setup. We introduce a Lagrangian density $\mathcal{L}$ made up of a scalar field $\phi$ and an abelian gauge field coupled to $G(\phi)$. The behavior of $G(\phi)$ coupled to the gauge field $F^{\mu\nu}F_{\mu\nu}$ produces strong interaction among the glueballs in a simpler manner \cite{Adamu, Adamu1, Brito}. The $G(\phi)$ coupled to the gauge field absorbs the long-distance dynamics of the photon propagator and prevents it from decoupling at longer wavelengths. %That ensures that the propagator does not decouple at longer wavelengths. 
We draw inspiration from the numerous arguments for an abelian dominance in QCD in the confining region. The abelian projection was initially proposed to justify the magnetic monopole condensation in the QCD vacuum, and infrared (IR) abelian dominance \cite{Hooft1, Ezawa} in QCD theory. Meanwhile, studies show that about $92\%$ of the QCD string tension is abelian \cite{Shiba}, while $\text{SU}(2)$ and $\text{SU}(3)$ QCD lattice simulations also point to abelian dominance \cite{Suzuki, Stack:2002kh, DIK:2003alb, Bali:1996dm, Sakumichi}. Other recent references can be found in \cite{Haresh, RAVAL2019545} for interested readers. 
Glueballs are potential \textit{dark matter candidates} with important cosmological and astrophysical implications, making them an intriguing subject for particle-physics investigations \cite{Boddy, Boddy:2014qxa, Dietrich, Appelquist}.

To investigate the cosmological phase transitions within this framework, we integrate out the gauge field and obtain an effective \textit{thermodynamic potential} $\Phi$ \cite{Lombardo, Schmitt, Laine}, which acts as a radiative correction to the SSB potential $V(\phi)$. In this case, an effective potential, $V_{eff}(\phi,T)$, is defined with a unique minimum $\langle\phi\rangle=0$ at high-temperature regions $T\geq T_C$, with $T$ an arbitrary temperature and $T_C$ a critical temperature. At this temperature region, all elementary particles become massless; hence, the initially broken symmetry is restored \cite{Kolb, Mukhanov}. 
This behavior is similar to the QCD running coupling, which decreases with increasing temperature and density due to asymptotic freedom. Consequently, it is possible to study the high baryon density regime of the early universe at temperatures up to $T\sim \text{M}_p\sim 10^{19}\text{GeV}$ and densities close to $\rho_p\sim \text{M}^4_p\sim 10^{94}g\text{cm}^{-3}$. Below $T_C$, the $V_{eff}(\phi,T)$ thermodynamically favors a degenerate vacuum, so a phase transition \cite{Ovrut, PhysRevLett.53.732, PhysRevD.30.2061, Ovrut:1984gb, Linde} is possible due to different kinds of interactions among the particles. It is noteworthy that the properties and the laws governing elementary particles change significantly below $T_C$. In QCD, particles are confined in this region, leading to the breakdown of perturbative QCD. That notwithstanding, modern-day elementary particle theory allows the description of super-dense nuclear matter \cite{Linde5}.

Furthermore, we investigate the phenomena of confinement, deconfinement, and cosmological phase transitions by analyzing the scalar potential $V(\phi)$ and its thermal counterpart $V_{eff}(\phi, T)$ separately. We aim to identify the regimes in the evolution of the universe in which glueballs exist in either confined or deconfined phases. Within this framework, we define a critical temperature $T_C$ with the following expectations: for $T < T_C$, glueballs are confined, characterized by a positive squared glueball mass $m^2(T)$ and a negative cosmological constant $\Lambda$. At the transition point $T = T_C$, we expect the onset of deconfinement, accompanied by vanishing $m^2(T)$ and $\Lambda$. For $T > T_C$, glueballs remain deconfined, exhibiting an unstable (negative) squared mass $m^2(T)$ and a positive $\Lambda$ driven by the cosmological phase transition.

The critical temperatures associated with weak and electromagnetic symmetry restoration occur at approximately $T_C \sim 10^2\,\mathrm{GeV} \sim 10^{15}\,\mathrm{K}$, while grand unified theories (GUT) predict a strong--electroweak symmetry--restoration scale near $T_C \sim 10^{15}\,\mathrm{GeV} \sim 10^{28}\,\mathrm{K}$. Such extreme temperatures are far beyond current experimental capabilities---considering that even the peak temperatures achieved in supernova explosions reach only $\sim 10^{11}\,\mathrm{K}$. Consequently, these high-energy phase transitions could only have occurred in the earliest stages of the universe. During this epoch, the strong, weak, and electromagnetic interactions were unified, and as the universe cooled, it underwent a sequence of symmetry-breaking phase transitions \cite{Kirzhnits, Kirzhnits1, Weinberg, Dolan, Kirzhnits-Linde, Linde7}. Finally, we employ the Mean Field (MF) approximation to study the formation of dense matter composed of glueballs and gluons---an approach commonly used to investigate the internal dynamics of compact astrophysical objects such as neutron stars, white dwarfs, quark stars, and hybrid stars~\cite{Menezes}.

The manuscript introduces a unified theoretical framework that directly links the nonperturbative confinement dynamics of QCD to the inflationary evolution of the early universe through the study of phase transitions. The central innovation of this work lies in the dual role played by a single scalar field $\phi$, which, when coupled to an effective abelian gauge field through a color dielectric function $G(\phi)$, simultaneously generates a Cornell-like confining potential relevant to hadronic physics and serves as the inflaton in cosmology. This mechanism intrinsically links the QCD gluon condensate to the cosmological constant, proposing that the deconfinement transition of strong-interaction matter is not merely an analogy but is the physical engine driving a major epoch in the early universe. The model, therefore, provides a concrete bridge between nonperturbative QCD dynamics and the thermal and cosmological evolution of the early universe.

In this paper, we introduce the Lagrangian density and the potential that form the basis of the study in Sec.~\ref{model} and also throw light on the QCD-like and cosmological properties of the model. In Sec.~\ref{CG} we broadly discuss the confinement of glueballs and calculate its phenomenological confining potential $V_c(r)$ in the form of a Cornell-like potential for confining heavy quarks. In Sec.~\ref{thermal} we present the procedure for computing thermal corrections to the effective potential. We proceeded to calculate the \textit{plasma mass} associated with it and show the first-order phase transition driven by the thermal correction to $V(\phi)$. In Sec.~\ref{CG1}, we use the MF approximation method to calculate the behavior of the nuclear matter formed by glueballs and gluons admixture. Here, we introduce a dimensionless glueball field $\chi\,=\,\phi/\phi_0$ to facilitate the study. Also, we study the thermal fluctuating gluon condensate, and the thermal fluctuating effective glueball mass, $m_\chi^*$. Sec.~\ref{cosmology} has four subsections where we discuss QCD-like and cosmological consequences of the model. In Sec.~\ref{cosmology1} we throw light on the behavior of the vacuum energy density of the universe and its implications. In Sec.~\ref{cosmology2} we discuss the domain wall problem of the potential and how to remove it for smooth phase transitions to occur. We also discuss the inflationary universe in Sec.~\ref{cosmology3} by studying various scenarios. In Sec.~\ref{analysis} we present the analysis and conclusion.  We use natural (Gaussian) units $\hbar\,=\,c\,=\,k_B\,=\,G_N=\,1$ and metric signature $\text{diag(1,\,-1,\,-1,\,-1)}$ throughout the paper.

\section{The model}\label{model}
We start with Lagrangian density given as:
\begin{equation}\label{1}
\mathcal{L}=\dfrac{1}{2}\partial_\mu\phi\partial^\mu\phi-\dfrac{1}{4}G(\phi)F^{\mu\nu}F_{\mu\nu}-V(\phi),
\end{equation}
where $G(\phi)$ is a dimensionless function commonly referred to as the color--dielectric function \cite{aIssifu, Brito}. As discussed in detail in Ref.~\cite{Issifu:2022pif}, $G(\phi)$ parametrizes the effective strong interaction between the scalar field $\phi$, which will be later identified as the glueball field. %{\color{red} TF: I think the interpretation as a glueball is misleading, we have to polish the nomenclature and name the excitation of $\phi$ in a different way. Much of my difficulty with the interpretation in this work is related to the association with a glueball.  It could be instead a collective excitation of  gluons, or correlated gluons, as an effective field}. 
Its primary role is to encode the nonperturbative, long-distance dynamics of the gauge field, most notably color confinement. In this framework, $G(\phi)$ acts as an effective, field-dependent gauge coupling: in the confining regime, one typically has $G(\phi)\!\to\!0$, which suppresses gluonic fluctuations and induces a large effective coupling, $g_{\rm eff}\propto G(\phi)^{-1/2}$. On the other hand, $G(\phi)\!\to\!1$, the theory approaches the perturbative phase where gluons propagate freely. Thus, $G(\phi)$ provides a phenomenological bridge between short-distance asymptotic freedom and long-distance confinement, allowing the model to capture the essential infrared features of QCD. 

We now introduce the SSB potential in a form 
\begin{equation}\label{2}
V(\phi)=\dfrac{\lambda\phi^4}{4}-\dfrac{m^2\phi^2}{2},
\end{equation}
where $\lambda$ is a dimensionless coupling constant and $m$ is the mass of the scalar field. The potential is invariant under discrete symmetry transformation $\phi\leftrightarrow -\phi$, whilst its minimum determined under the conditions $V'(\phi)=0$ and $V''(\phi)>0$ are $\langle\phi\rangle=\phi_0=\pm m/\sqrt{\lambda}$. Also, 
\begin{equation}\label{p1a}
V(\phi_0)=-\dfrac{m^4}{4\lambda} \qquad{\text{and}}\qquad V''(\phi_0)=2m^2. 
\end{equation}
From the QCD trace anomaly, the energy-momentum tensor trace $\Theta^\mu_\mu$ is nonvanishing 
\begin{equation}
   \left\langle \Theta^\mu_\mu\right\rangle=-\dfrac{9}{8}\left\langle\dfrac{\alpha_s}{\pi}G_{\mu\nu}G^{\mu\nu}\right\rangle,
\end{equation}
where $\alpha_s$ is the strong coupling constant and $G^{\mu\nu}$ is the non-abelian gauge field strength. Using parton-hadron duality principle \cite{Schechter,Carter,Campbell,Kochelev1},
\begin{equation}\label{ph}
    \left\langle \Theta^\mu_\mu\right\rangle=\left(4V(\phi)-\phi\dfrac{dV(\phi)}{d\phi}\right)\Big\vert_{\phi_0}=-\dfrac{m^4}{\lambda},
\end{equation}
 we can identify the Bag constant $-B_0\,=\,-m^4/\lambda$. Relating the gluon condensate to the Bag constant, we can identify the coupling constant
 \begin{equation}\label{csc}
     \lambda=\dfrac{8m^4\pi}{9\left\langle\alpha_sG_{\mu\nu}G^{\mu\nu}\right\rangle}. %=\dfrac{2m_\phi^4\pi}{9\left\langle\alpha_sG_{\mu\nu}G^{\mu\nu}\right\rangle}.
 \end{equation}
  Since the coupling $\lambda$ is related to the gluon condensate $\left\langle G_{\mu\nu}G^{\mu\nu}\right\rangle$, we can identify $m^2$ with the \textit{glueball mass} since glueballs have gluon degrees of freedom.

%-----------------------------------------------------------------------------------------------------------------------

 Additionally, $V(\phi)$ has a `false' vacuum at $\phi_0=0$ (this point represents the maxima of $V(\phi)$). However, quantum theory should be developed around a `true' vacuum of the corresponding classical potential. Hence, the $V(\phi)$ has a degenerate vacuum that breaks its initial discrete symmetry. 

%-------------------------------------------------------------------------------------------------

\section{Confinement Of Glueballs}\label{CG}

We choose a \textit{color dielectric function} defined as
\begin{equation}\label{3}
G(\phi)={2\kappa^2\phi^2};
\end{equation}
with $\kappa$ the decay constant of the glueballs with dimension, $[\kappa]\,=\,\text{GeV}^{-1}$.  The $G(\phi)$ is carefully defined to ensure that we obtain the needed QCD-like properties (confinement and asymptotic freedom behavior) of the glueballs and the cosmological phase transitions consistent with the objectives of the study. {It is worth noting that the choice of $G(\phi)$ is open; it can be any polynomial in the order of $\phi^{\tilde{n}}$, with $\tilde{n}\in\mathbb{N} $, depending on the objectives of the investigator. } 

The equations of motion of the Lagrangian Eq.(\ref{1}) are;
\begin{equation}\label{4}
\partial_\mu\partial^\mu\phi+\dfrac{1}{4}\dfrac{\partial G}{\partial\phi}F^{\mu\nu}F_{\mu\nu}+\dfrac{\partial V}{\partial\phi}=0
\end{equation}
and 
\begin{equation}\label{5}
\partial_\mu[G(\phi)F^{\mu\nu}]=0.
\end{equation}
We solve the above equations from the {static sector of $\phi$} using spherical coordinates i.e. $\mu=1,2,3$ and also choose the appropriate indices for the gauge field strength such that only the \textit{chromoelectric flux} is available for the analysis. {Since we are interested in investigating \textit{color confinement} of static color particles (valence gluons), the electric field is sufficient for the analysis, hence $\nu=0$. %hence $j^\nu=(\rho,\,\vec{0})$, with $\nu=0$, $j^\nu$ is the four-current and $\rho$ is the charge density. 
This choice removes the effect of spin and magnetic field from the analysis. As a result,}
\begin{align}\label{6a}
&\dfrac{1}{r}\dfrac{d}{dr}[r^2G(\phi){E}]=0\rightarrow\nonumber\\
&E=\dfrac{\tilde{\Lambda}}{r^2G(\phi)},
\end{align}
where $\tilde{\Lambda}=q/4\pi\varepsilon_0$, is the integration constant, $q$ an electric charge, $\varepsilon_0$ is the permittivity of free space, and $E$ the electric field. {It is important to mention that even in the absence of a source, Maxwell's equations in free space have static spherically symmetric solutions. For interested readers, this phenomenon has been applied in studying charged strings in Ref.\cite{ Cvetic} and confinement in Ref.\cite{Brito}}. Thus Eq.(\ref{4}) becomes 
\begin{align}\label{5a}
\dfrac{d^2\phi}{dr^2}+\dfrac{2}{r}\dfrac{d\phi}{dr}+2\kappa^2\phi\left(\dfrac{\tilde{\Lambda}}{r^2G(\phi)}\right)^2-V'&=0\nonumber\\ 
\phi''+\dfrac{2}{r}\phi'+m^2\phi&=0;
\end{align}
we substituted $F^{\mu\nu}F_{\mu\nu}=F^{j0}F_{j0}+F^{0i}F_{0i}=-2E^2$ also, $\phi$ prime represents derivative with respect to $r$ and double prime is the second derivative. In the confinement regime, the inter-particle separation $r$ is large compared to the glueball Compton wavelength ($r \gg m^{-1}$), the scalar field $\phi$ varies slowly and remains close to its vacuum expectation value, rendering $G(\phi)=2\kappa^{2}\phi^{2}$ approximately constant. Consequently, the term $1/[r^{4}G(\phi)^{2}]$ decays as $1/r^{4}$, faster than the $1/r^{2}$ scaling of the derivative terms, and can therefore be neglected at leading order. 
One of the physically motivated solutions to the above equation is 
\begin{equation}\label{5b}
\phi(r)=\dfrac{\sin(m r)}{r}.
\end{equation}
We chose this solution because it enables us to visualize both \textit{color confinement} and \textit{asymptotic freedom} in the framework of Cornell-like confining potential.
Using the well-known relation for determining the electrodynamic potential 
\begin{equation}\label{5c}
V(r)=\int{E dr},
\end{equation}
and substituting Eq.(\ref{6a}), we obtain
\begin{align}\label{5d}
V_c(r)&=\int{\dfrac{\tilde{\Lambda}}{2\kappa^2\left({\sin(mr)}\right)^2}}\nonumber\\
&=-\dfrac{\tilde{\Lambda}\cot(mr)}{2m\kappa^2}+k \nonumber\\
&= -\dfrac{q}{4\pi\varepsilon_0}\dfrac{\cot(mr)}{2m\kappa^2},
\end{align}
where $k$ is the integration constant representing the self-energy of the glueballs. It has been set to $k=0$ in the last step.

\begin{figure}[t]
  \centering
 \includegraphics[width=0.5\linewidth]{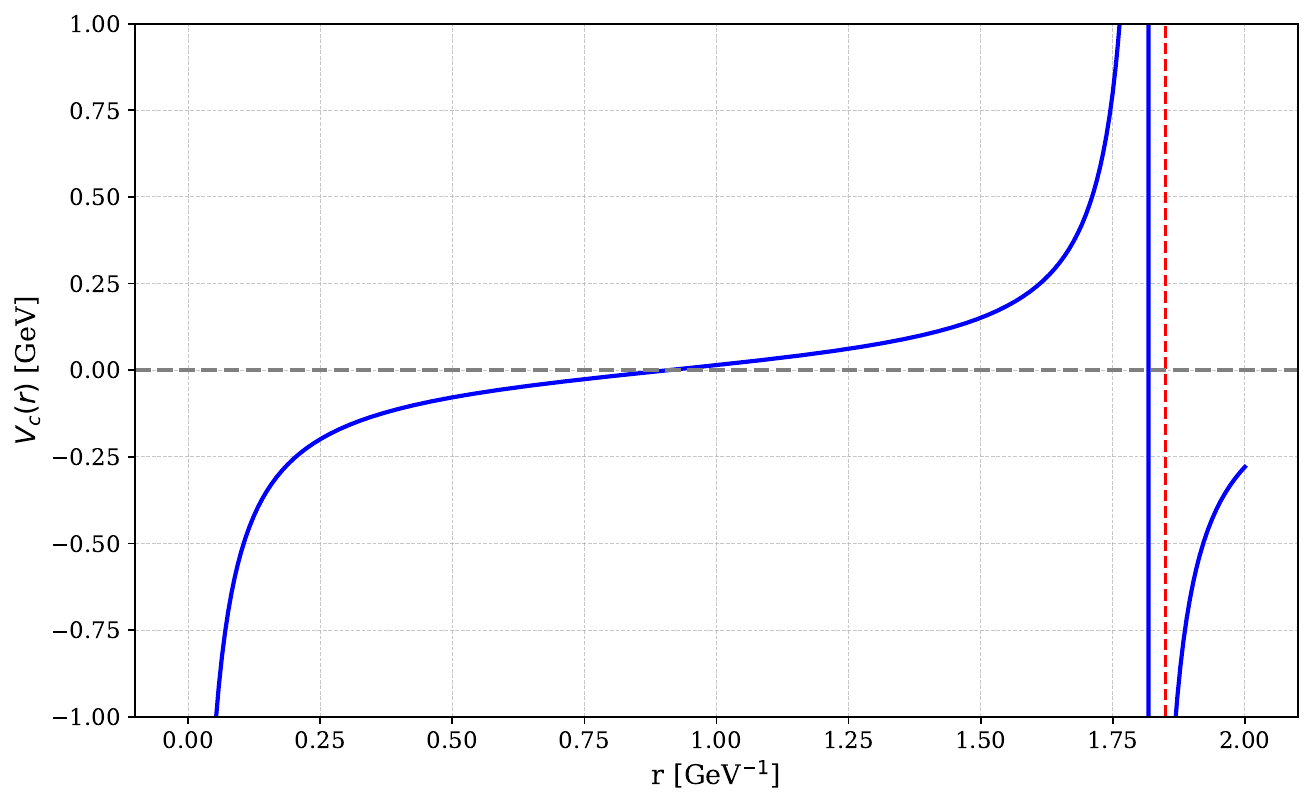}
   \caption{A graph of $V_c(r)$ against $r$. The graph shows a linear rise in $V_c(r)$ with $r$ indicating confinement of the glueballs with increasing $r$, below $r=0.9\,GeV^{-1}$ the glueballs are asymptotically free. When we increase $r$ beyond $\sim\,1.85\,{GeV^{-1}}$, there is hadronization represented by a vertical red line in the graph \cite{conell:1, Buisseret}. That is the point at which the string tension that holds the particles in a confined state breaks, leading to the formation of mesons.}
   %\floatfoot{}
   \label{pba}
\end{figure}
The choice of the glueball decay parameter $\kappa = 0.5~\mathrm{GeV}^{-1}$ follows from matching the linear part of the confining potential in Eq.~(\ref{5d}) to the known QCD string tension. Since Eq.~(\ref{5d}) becomes linear when $\cot(mr)\simeq -mr$, the linear part of the potential reduces to $V_c(r)\approx (q/8\pi\varepsilon_0)\,r/\kappa^2$, giving a string tension $\sigma_{\rm string}=q/(8\pi\varepsilon_0\kappa^2)$. Using the standard lattice value $\sigma_{\rm QCD}\approx 0.18\!-\!0.20~\mathrm{GeV}^2$ \cite{conell:1, Buisseret} and $q=\varepsilon_0=1$ (natural units) yields $\kappa\approx 0.45~\mathrm{GeV}^{-1}$. Rounding up to the nearby value $\kappa=0.5~\mathrm{GeV}^{-1}$ provides a reasonable choice that reproduces the correct confinement scale and yields the Cornell-like behavior in Fig.~\ref{pba}, consistent with lattice QCD \cite{Shiba, Bali:2000gf, Athenodorou:2021qvs}.

The vertical red line in the diagram represents the threshold at which the particles begin to hadronize into light hadrons \cite{Casher, Smith, Collinsj, Webber, He} due to QCD string breaking. The use of an effective Abelian gauge field $A_\mu$ coupled to a color dielectric function  $G(\phi)$ provides a well-established simplification of nonperturbative QCD, motivated by Abelian dominance observed in lattice simulations \cite{Hooft1, Suzuki:1989gp}. In this approach, the non-Abelian long-distance dynamics are encoded in the scalar glueball field through $G(\phi)$, which acts as a medium-dependent dielectric function that vanishes in the confining vacuum and suppresses the chromoelectric field \cite{Issifu:2022pif}. Equation (\ref{6a}) implies a static solution $E \propto 1/(r^{2} G(\phi))$, so as $G(\phi) \to 0$ the field is screened and its energy grows linearly with separation, yielding a confining Cornell-like potential \cite{Brito}. This minimal setup effectively captures the formation of a chromoelectric flux 
tube while remaining tractable for studies of thermal and cosmological phase transitions.

\section{Thermal Correction to the Potential and Phase Transitions}\label{thermal}
From the Lagrangian in Eq.(\ref{1}), we can integrate out the gauge field using path integral formalism
\begin{equation}\label{22}
Z[A]=\int{D[A]e^{iS[A]}}.
\end{equation}
The intention is to convert the gauge field $F^{\mu\nu}F_{\mu\nu}$ into a thermodynamic potential $\Phi$ so we can study its contribution to phase transitions in an $V_{{eff}}(\phi, T)$. 
The result obtained here, together with the $G(\phi)$, will serve as a thermal correction to the $V(\phi)$. In that case, we can obtain an effective potential, relevant for analyzing the {\it cosmological phase transitions}. Additionally, the {\it domain wall} problem associated with the SSB model will be removed at higher temperatures. Hence, at higher temperatures, the initially broken-down symmetry gets restored, and the $V_{{eff}}(\phi, T)$ will have a unique minimum, $\langle\phi\rangle=0$. 

From Eq.(\ref{1}), we can write the action for the gauge field as 
\begin{align}\label{23}
S[A]&=\dfrac{1}{2}\int{d^4x A^\nu\left[\eta_{\mu\nu}\partial^2-\partial_\mu\partial_\nu\right]A^\mu}\nonumber\\
&=\dfrac{1}{2}\int{d^4x A^\nu[\eta_{\mu\nu}\partial^2-\left(1-\dfrac{1}{\alpha}\right)\partial_\mu\partial_\nu]A^\mu},
\end{align}
we expressed $F_{\mu\nu}=\partial_\mu A_\nu-\partial_\nu A_\mu$, and in the last step, we have introduced the gauge fixing term $({1}/{2\alpha})(\partial_\mu A^\mu)^2$ \cite{Lombardo}. Using the Euclidean transformation 
\begin{equation}\label{24}
x=(it,\bar{x}), \qquad{}\qquad t\rightarrow -i\tau,
\end{equation}
with $t$ the Euclidean time. We obtain the partition function
\begin{align}\label{25}
Z[A]&=\int{D[A]\exp\left(\dfrac{1}{2}\int_X{A^\nu[\eta_{\mu\nu}\partial^2-(1-1/\alpha)\partial\mu\partial_\nu]A^\mu} \right) }\nonumber\\
&=\int{D[A]\exp\left(\dfrac{1}{2}\int_X{A^\nu\partial^2A_\nu} \right)},
\end{align}
here, 
{\begin{equation}
    \int_X\equiv\int_0^\beta d\tau\int d^3x.
\end{equation}}
We used the Feynman gauge $\alpha=1$ to arrive at the last expression in Eq.(\ref{25}).
We can now introduce the Fourier transform of the thermal fluctuating gauge fields 
\begin{equation}\label{26}
A^\nu(x)=\dfrac{1}{\sqrt{TV}}\sum_k e^{ikx}A^\nu(k),
\end{equation} 
with four-momentum defined as $k\equiv (k_0,\vec{k})=(-i\omega_n,\vec{k})$ therefore, $kx=k_0x_0-\vec{k}\cdot \vec{x}=-(\tau \omega_n+\vec{k}\cdot\vec{x})$ \cite{Schmitt}. The gauge field follows the periodicity condition 
\begin{equation}\label{26a}
A_\nu(0,x)=A_\nu(\beta,x)\quad{:}\quad e^{i\omega_n\beta}=1 \quad{\rightarrow}\quad \omega_n=2n\pi T \quad{\text{where}}\quad n=0,1,2\cdots,
\end{equation}
with $\beta=1/T$. So, Eq.(\ref{25}) becomes 
\begin{equation}\label{27}
Z[A]=\int{D[A]\exp\left( \dfrac{1}{2T^2}\sum_k A^\nu(k)\partial^2A_\nu(k)\right)},
\end{equation}
%{\color{red}TF: $\partial^2\to k^2$???} {\color{blue}AI: We used $k_\mu\rightarrow i\partial_\mu$ defined below Eq.~(25).}
{we have substituted  
\begin{equation}
    \int_Xe^{ikx}=\dfrac{V}{T}\delta_{k,0} .
\end{equation}}
 Using the general relation 
\begin{equation}\label{28}
\int{\prod_k^N d\eta_k^\dagger d\eta_k \exp\left(-\sum_{i,j}^N\eta^\dagger_i D_{ij}\eta_j \right) }=\det D,
\end{equation}
we can express Eq.(\ref{27}) as
\begin{equation}\label{29}
Z=\det\left(-\dfrac{\partial^2}{2T^2}\right).
\end{equation}
Using the quantum operator form of the four-momentum, $k_\mu\rightarrow i\partial_\mu$, 
and $k^2=(\omega_n^2+\vec{k}^2)$ yields
\begin{equation}\label{30}
Z=\det\left(\dfrac{k^2}{2T^2}\right)=\prod_{n,k}\left[\dfrac{\omega_n^2+\vec{k}^2}{2T^2} \right]^2, 
\end{equation} 
By using the dispersion relation 
\begin{equation}\label{30a}
E_k^2=\vec{k}^2+m^2_k,
\end{equation}
and noting that $m_k^2=0$ for gluon and photon fields, Eq.(\ref{30}) reads
\begin{equation}\label{31}
\ln Z=2\sum_{n,k}\ln\left(\dfrac{\omega_n^2+E_k^2}{T^2} \right). 
\end{equation}
Performing the sum over the Matsubara frequency $\omega_n$ yields,
\begin{equation}\label{32}
\sum_{n}\ln\left( \dfrac{\omega_n^2+E^2_k}{T^2}\right)=\dfrac{E_k}{T}+2\ln\left(1-e^{-E_k/T}\right)+const.
\end{equation}
Therefore, using the above result together with Eq.(\ref{31}) yields the thermodynamic potential, $\Omega=-T\ln Z$ \cite{Lombardo, Schmitt, Laine}
\begin{equation}\label{33}
\Phi=\dfrac{\Omega}{V}=-2T\int{\dfrac{d^3k}{(2\pi)^3}\left[\beta E_k+2\ln\left(1-e^{-E_k/T}\right) \right]},
\end{equation}
{where we have changed the summation in $k$ into an integral in the above expression
\begin{equation}
    \sum_k\rightarrow\int\dfrac{d^3k}{(2\pi)^3}.
\end{equation}}
Subtracting the ground state energy $\beta E_k$ and resolving the remaining equation, yields
\begin{align}\label{34}
\Phi&=-\dfrac{T}{\pi^2}\int{k^2dk \ln\left(1-e^{-E_k/T}\right) }\nonumber\\
&=-\dfrac{2}{3\pi^2}\int_0^\infty{\dfrac{k^3dk}{e^{-E_k/T}-1}}\nonumber\\
&=-\dfrac{2T^4}{3\pi^2}\int_0^\infty{ \dfrac{x^3dx}{e^{x}-1}}\nonumber\\
&=-\dfrac{2\pi^2T^4}{45}.
\end{align}
We used integration by parts to arrive at the second step while in the third step, we parameterize $x=-E_k/T$. Hence, the Lagrangian in Eq.(\ref{1}) becomes 
\begin{equation}\label{35}
\mathcal{L}_{eff}=\dfrac{1}{2}\partial_\mu\phi\partial^\mu\phi-V_{eff}(\phi,T),
\end{equation}
where 
\begin{equation}\label{36}
V_{eff}(\phi,T)=\dfrac{\lambda\phi^4}{4}-\dfrac{m^2\phi^2}{2}+\dfrac{4\pi^2(\kappa T^2)^2\phi^2}{45}.
\end{equation}
{The last term in Eq.~(\ref{36}) originates from the $G(\phi) F^{\mu\nu} F_{\mu\nu}$ contribution in Eq.~(\ref{1}), where $F^{\mu\nu} F_{\mu\nu}$ has been integrated out and incorporated as a thermal correction to the original potential $V(\phi)$ in Eq.~(\ref{2}) coupled with $G(\phi)$.} 
%
%{\color{red} TF: it will help including the formula that connects $\Phi$ and the thermal average  with the thermal correction to $V_{eff}$ equal to $-\frac14G(\phi)\langle F^{\mu\nu}F_{\mu\nu}\rangle_{T}$.} {\color{blue} This is precisely the third therm in Eq.~(\ref{36}).}
%
This temperature-dependent contribution plays the role of a quantum correction to the classical potential. Quantum corrections are typically incorporated by evaluating all one-particle-irreducible vacuum diagrams in the Lagrangian formalism, implemented through the shift $\mathcal{L}(\phi + \phi_T)$, where $\phi_T$ represents the thermal correction. Terms linear in $\phi$ are neglected in this expansion, as discussed in Refs.~\cite{Coleman, Jackiw}. We adopt the same reasoning here. Thus, the last term in Eq.~(\ref{36}) acts as a temperature correction to $V(\phi)$, shifting the vacuum energy density due to quantum fluctuations. A detailed discussion of how to introduce temperature-dependent quantum corrections into an effective classical potential is provided in Refs.~\cite{Linde5, Collins}.

This procedure amounts to shifting the vacuum according to $\phi \rightarrow \phi + \phi_T$, allowing the temperature-induced shift to be identified from Eq.~(\ref{36}) as %{\color{red}
\[
\phi_T =\pm\,\sqrt{ \frac{m^2}{\lambda} -\frac{8\pi^2 \kappa^2 T^4}{45\lambda}}.
\]
It is important to note that thermal corrections introduced through the scalar field itself typically scale as $T^2$~\cite{Linde5, Collins}, whereas the correction arising through the gauge field in our framework scales as $T^4$. In both cases, however, the modification appears in the quadratic term of the potential, and the different powers of $T$ are dimensionally consistent.

Witten~\cite{Witten} discussed in detail how temperature corrections to symmetry-breaking potentials can describe the transition from a high-temperature early universe with a unique vacuum at $\phi_0 = 0$ to a low-temperature phase $T < T_C$ with degenerate vacua. Ref.~\cite{Witten} also shows that ``supercooling'' can occur at $T = T_C$, where the system remains trapped in the false vacuum at $\phi_0 = 0$ due to tunneling through potential barriers. As we shall see below,  in our model, we observe that the temperature-dependent term decouples at $T = T_{c\phi}$, leaving only the $\lambda \phi^4$ term before the temperature dependence reappears for $T > T_{c\phi}$. Just below $T_{c\phi}$, the thermodynamically favored configuration is the degenerate vacuum; however, the transition to this state requires quantum tunneling through a barrier, which proceeds slowly. 
Consequently, the system undergoes supercooling in the false vacuum $\phi_0 = 0$ until the temperature decreases sufficiently for the tunneling rate to become significant. We define the critical temperature $T_{c\phi}$ as
\begin{equation}\label{37}
T_{c\phi}=\left(\dfrac{45 m^2}{8\pi^2\kappa^2}\right)^{1/4},
\end{equation} 
hence,
\begin{equation}\label{39}
V_{eff}(\phi,T)=\dfrac{\lambda\phi^4}{4}-\dfrac{m^2(T)\phi^2}{2},
\end{equation}
with 
\begin{equation}\label{38}
m^2(T)={m^2}\left[1-\dfrac{T^4}{T_{c\phi}^4} \right].
\end{equation}
Consequently, the thermal fluctuating glueball mass $m(T)$ decreases with temperature and vanishes at $T=T_{c\phi}$. See Fig.~\ref{2}. The use of a temperature-dependent glueball mass that becomes tachyonic above $T_{c\phi}$ is well motivated by effective potential analyses of gauge theories \cite{Pisarski:1983ms, Megias:2004hj} and lattice QCD results \cite{Boyd:1996ex}, which shows that the gluon condensate melts and the confined vacuum becomes unstable near $T_{c\phi}$. This temperature dependence provides a smooth interpolation through the (de)confinement transition and naturally drives the system toward the symmetric minimum at $\phi = 0$, at $T\geq T_{c\phi}$, in agreement with glueball effective theory \cite{Arikawa:2025kjx} and phenomenological/lattice-based studies \cite{Buisseret:2010mop}. This negative-squared glueball mass phenomenon, which occurs under extreme conditions of temperature, has also been noted in studies of the quark–gluon plasma \cite{Kochelev2, Mathieu}.

The $V(\phi)$ produces an inflationary scenario in a simple form \cite{Linde1} with reasonable restrictions. 
It has been established that during the evolution of the early universe, phase transitions occurred due to the spontaneous breakdown of gauge symmetries. Therefore, transitioning from the hot Big Bang to inflation, the scalar field $\phi$ that describes the system requires coupling to `matter'. Such coupling introduces the necessary correction to the potential, giving it the expected behavior after inflation (an example of such correction to one loop can be found in Ref.\cite{Coleman}).

At $T=0$ the $V_{eff}(\phi,T)$ has a negative square mass determined at $\phi_0=0$. The imaginary mass generates an exponentially growing solution where $\phi$ grows until it finds its true ground state. This challenge is solved when we place $\phi$ in contact with a thermal bath. So, there is an interaction between $\phi$ and the particles in the thermal bath that dampens the exponential growth. This procedure introduces a {\it plasma mass} 
\begin{equation}
m^2_{plasma}=\dfrac{8\pi^2}{45}(\kappa T^2)^2,
\end{equation}
where ($\kappa T^2$) is precisely the plasma temperature \cite{Kolb}, this expression is determined from Eq.(\ref{39}) with the full definition of $T_{c\phi}$ substituted. So, at a finite temperature, the \textit{effective mass $m_T$} at the classical vacuum $\phi_0 =0$ becomes
\begin{equation}\label{p1}
m^2_T=-m^2+m^2_{plasma}.
\end{equation} 
At temperatures where $m_T>0$, $\phi_0=0$ will be an unstable minimum of the $V_{eff}(\phi,T)$ resulting into SSB and when $m_T\leq 0$ at a particular temperature, then $\phi_0 =0$ becomes a stable minimum because the effective mass is real \cite{Kolb}. The behaviour of $m_T$ is controlled by $T_{c\phi}$, so at $T=T_{c\phi}$, $m_T=0$, $\phi_0 =0$ is a stable vacuum, at $T<T_{c\phi}$, $m_T>0$ and $\phi_0 =0$ is an unstable vacuum leading to SSB and at $T>T_{c\phi}$, $m_T<0$ and $\phi_0 =0$ is a stable vacuum. Therefore, the phase transition begins at $T=T_{c\phi}$, a point at which the \textit{effective potential} $V_{eff}(T,\phi)$ has a unique minimum. 

\begin{figure}[!t]
  \centering
 {\includegraphics[scale=0.5]{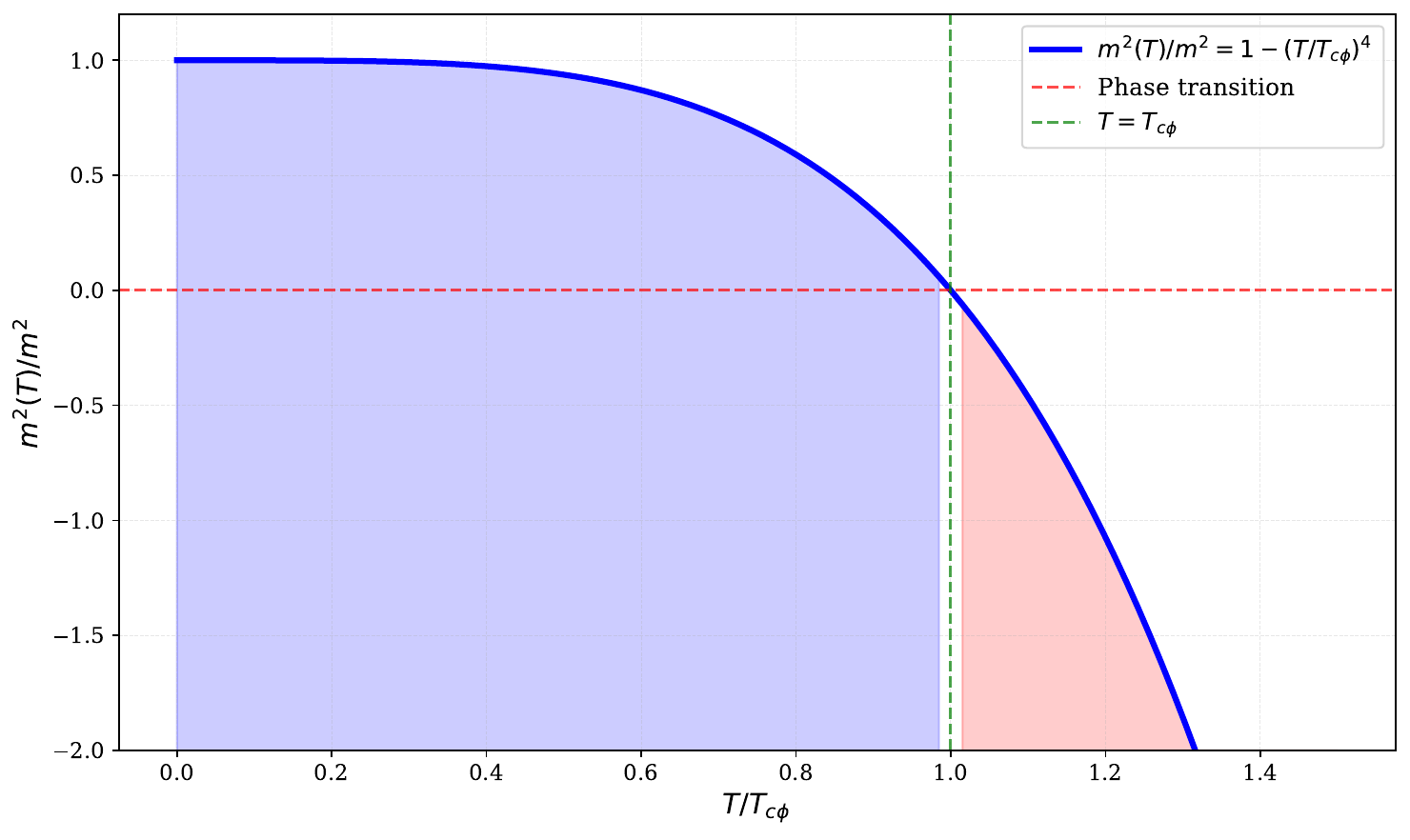}}
 \caption{A graph of $m^2(T)/m^2$ against $T/T_{c\phi}$. The $m^2(T)$ decreases rapidly with temperature and vanishes at $T=T_{c\phi}$, where a phase transition to a deconfined matter state is expected.}
   \label{pbb}
\end{figure}

\begin{figure}[!t]
  \centering
 {\includegraphics[scale=0.65]{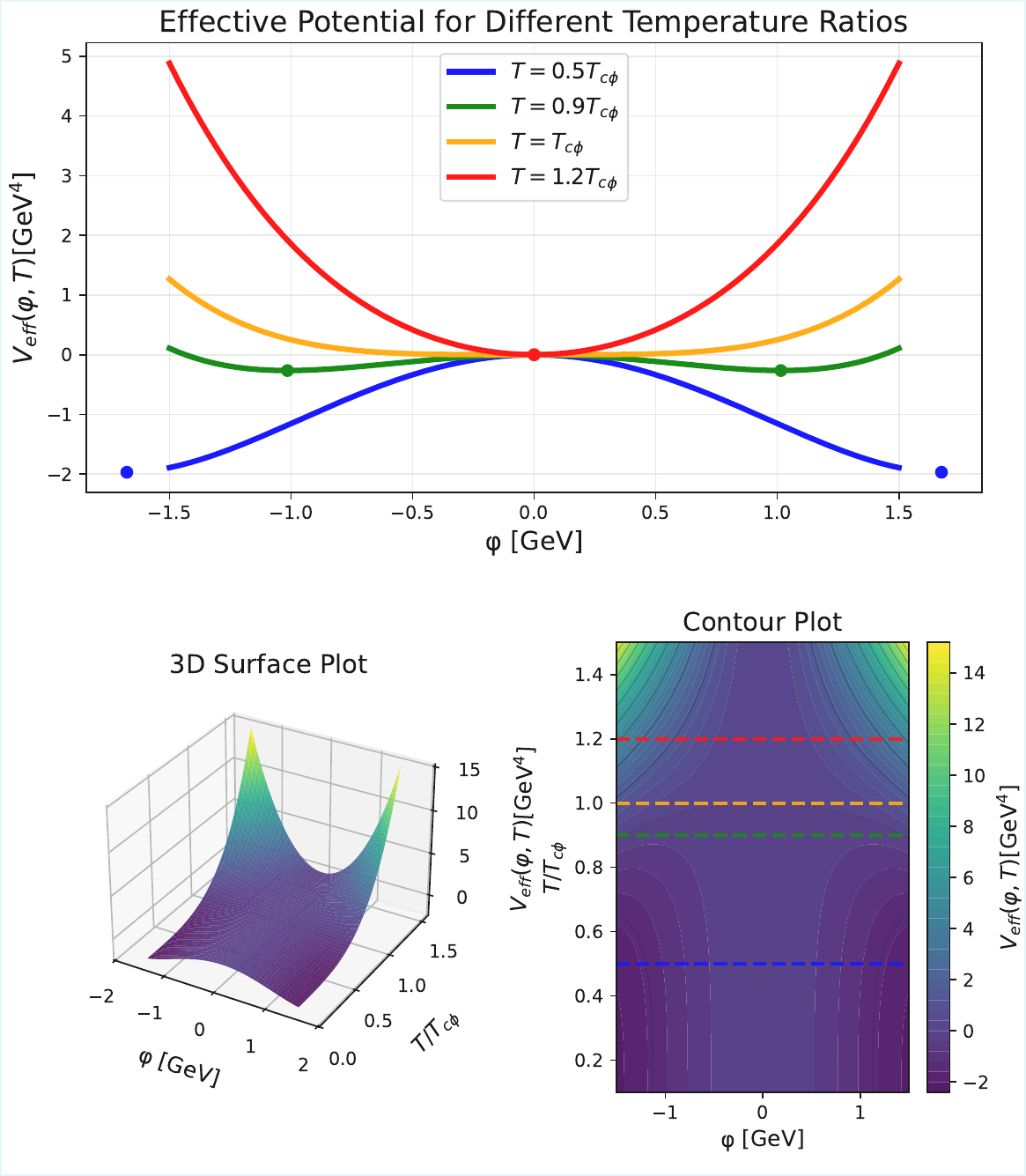}}
 \caption{A graph of $V(\phi,T)$ against $\phi,T$ for fix values of $T$. The graph shows a first-order phase transition at $T=T_{c\phi}$, at this point, inflation and cosmological phase transitions can be explored.}
   \label{sp}
\end{figure}
Figure~\ref{sp} illustrates the temperature dependence of the effective scalar potential $V_{eff}(\phi, T)$, including thermal corrections from the gauge field, and clearly reveals a first-order phase transition at the critical temperature $T = T_{c\phi}$ \cite{Ovrut, PhysRevLett.53.732, PhysRevD.30.2061, Ovrut:1984gb, Linde}. At high temperatures ($T>T_{c\phi}$), the potential exhibits a single minimum at $\phi = 0$, corresponding to the symmetric (deconfined) phase where the initially broken symmetry is restored \cite{Kolb, Mukhanov}. As the temperature drops below $T_{c\phi}$, a second, degenerate minimum appears (shown with bullet points in the plot), signaling spontaneous symmetry breaking and the onset of glueball confinement. The abrupt restructuring of the vacuum, marked by a pronounced energy barrier, is characteristic of the supercooling phenomenon discussed in \cite{Witten}, and provides the thermodynamic basis for the subsequent inflationary dynamics and glueball (de)confinement processes explored in this work.

\section{Dense Nuclear Matter}\label{CG1}
\begin{figure}[!t]
  \centering
 {\includegraphics[scale=0.5]{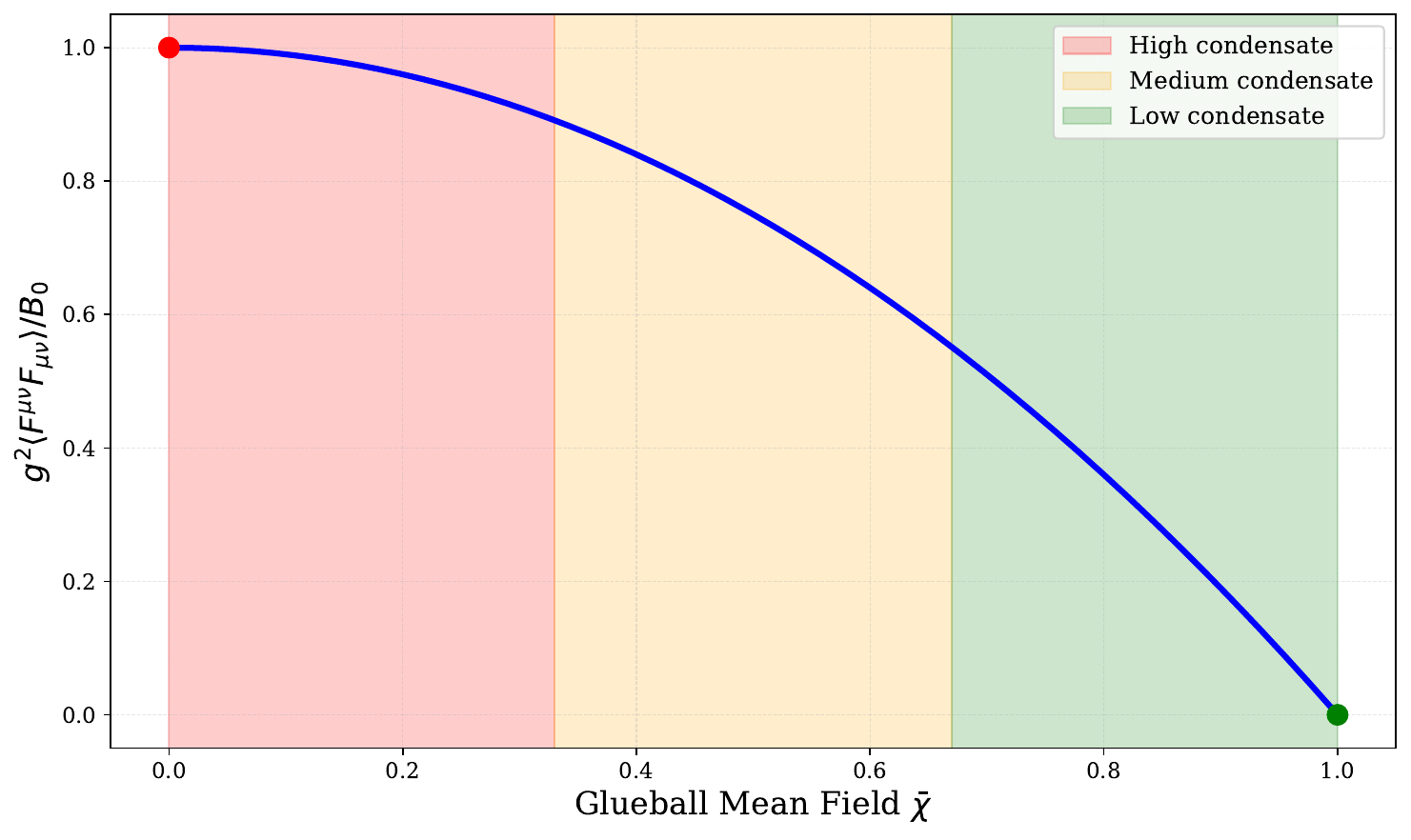}}
 \caption{A graph of $g^2\langle F^{\mu\nu}F_{\mu\nu}\rangle/B_0$ against $\bar{\chi}$. This graph shows that $g^2\langle F^{\mu\nu}F_{\mu\nu}\rangle/B_0$ decreases with increasing glueball mean field $\bar{\chi}$. We have a maximum condensate at $\bar{\chi}=0$ and a minimum condensate at $\bar{\chi}=1$.}
   \label{pc}
\end{figure}

In this section, we employ the MF approximation commonly used in studies of high-density nuclear matter \cite{Serot, Serot1}, following an approach analogous to the Walecka model for quantum hydrodynamics \cite{Walecka}. We apply this method to investigate strongly interacting matter composed of a gluon–glueball admixture. This phenomenological framework has proven successful in many contexts and can be extrapolated into the high-temperature and high-density regimes where nuclear equations of state (EoS) remain applicable. Our analysis focuses on the role of mean-field glueballs, the thermal fluctuations of the gluon condensate, and the resulting \textit{effective glueball mass}. To this end, we introduce a dimensionless, color-neutral glueball field, following the methodology outlined in Ref.~\cite{Carter}.
Substituting the definitions of the $V(\phi)$ and the $G(\phi)$ into Eq.~(\ref{4}) we get,
\begin{equation}\label{6}
\partial_\mu\partial^\mu\phi+(\kappa^2\phi)F^{\mu\nu}F_{\mu\nu}+(\lambda\phi^3-m^2\phi)=0.
\end{equation}
We define a dimensionless glueball field of the form $\chi=\phi/\phi_0$, which yields
\begin{equation}\label{7}
\phi_0\partial_\mu\partial^\mu\chi+(\kappa^2\phi_0)\chi F^{\mu\nu}F_{\mu\nu}+(\lambda\phi^3_0\chi^3-m^2\phi_0\chi)=0.
\end{equation}
Using Eq.~(\ref{ph}) we can identify the Bag constant $B_0=m^2\phi_0^2$, (we substitute the vacuum state $\phi_0=\pm m/\sqrt{\lambda}\rightarrow \lambda=m^2/\phi_0^2$) making a substitution in terms of $B_0$ into Eq.~(\ref{7}) yields,
\begin{equation}\label{10}
\phi_0^2\partial_\mu\partial^\mu\chi+(\kappa^2\phi^2_0)\chi F^{\mu\nu}F_{\mu\nu}-(B_0\chi-\lambda\phi^4_0\chi^3)=0.
\end{equation}
Again, using the vacuum state to eliminate $m$ from the definition of $B_0$, we get $\lambda\phi_0^4=B_0$ and
\begin{align}\label{11}
\phi_0^2\partial_\mu\partial^\mu\chi+g^2\chi F^{\mu\nu}F_{\mu\nu}-B_0(\chi-\chi^3)=0,
\end{align}
with $g^2=\kappa^2\phi_0^2$ a dimensionless constant. Using the MF approximation scheme, $\chi\rightarrow \langle\chi\rangle=\bar{\chi}$ and $F_{\mu\nu}\rightarrow\langle F_{\mu\nu}\rangle$, therefore
\begin{align}\label{12}
g^2\bar{\chi}\langle F^{\mu\nu}F_{\mu\nu}\rangle&=B_0\bar{\chi}\langle 1-\bar{\chi}^2\rangle\rightarrow\nonumber\\
g^2\langle F^{\mu\nu}F_{\mu\nu}\rangle&=B_0\langle 1-\bar{\chi}^2\rangle.
\end{align}
{We observed that $\bar{\chi}=0$ is always a trivial solution. For $\bar{\chi}\neq 0$, $\bar{\chi} = 1$ corresponds to a vanishing gluon condensate, whereas a nontrivial zero of $\bar{\chi}$ occurs when $g^2\langle F^{\mu\nu}F_{\mu\nu}\rangle=B_0$ (see Fig.~\ref{pc}). In the general case, the mean-field value of the glueball field is determined by the gluon condensate through
\begin{equation}
    \langle\bar{\chi}\rangle = \sqrt{1-\dfrac{g^2}{B_0}\langle F^{\mu\nu}F_{\mu\nu}\rangle},
\end{equation}
where we select the positive root, consistent with the physical requirement that the glueball amplitude be non-negative.
} %{\color{red} $\bar{\chi}=1$ TF: this is not a solution of Eq. 43 for static and homogeneous fields!} are solutions to the above expression. This is a condensate comprising both gluons and glueballs represented by $\langle F^{\mu\nu}F_{\mu\nu}\rangle$ and $B_0$ respectively. {\color{red} TF: the solution of Eq. 44 is $ \langle \bar \chi\rangle= 1- \frac{g^2}{B_0}\langle F^{\mu\nu}F_{\mu\nu}\rangle$, why  disregard the last term? Indeed in Eq. 47 and 49 this last term is nonzero!}

{Now, we define the glueball field as a sum of its mean field, $\bar{\chi}$ and a thermal fluctuating term, $\Delta$ i.e., $\chi=\bar{\chi}+\Delta$, imposing the restriction, $\langle\Delta\rangle=0$. This approach is used to introduce temperature into the condensate.} The dispersion relation for gluons and glueball excitations are
\begin{equation}\label{13}
E_A^2=k^2+m_A^{*2}\qquad{\text{and}}\qquad E_\chi^2=k^2+m_\chi^{*2}\,,
\end{equation}
where $m_A^*$ \cite{Silva:2016msq, Castorina:2007qv} and $m_\chi^*$ are gluons and scalar glueball masses with thermal fluctuations, respectively. %{\color{red} TF: include a reference for the gluon mass with thermal fluctuations. Is that known from Lattice QCD?}. 

We can compute $m^*_\chi$ directly from the Lagrangian density, obtained from Eqs.~\eqref{1}, 
\eqref{2} and \eqref{3} by expanding around the mean field, through 
\begin{equation}\label{14}
\phi^2_0 m^{*2}_\chi=-\left\langle \dfrac{\partial^2\mathcal{L}}{\partial\chi^2}\right\rangle =B_0[3(\bar{\chi}^2+\langle\Delta^2\rangle)-1]+{g^2}\langle F^{\mu\nu}F_{\mu\nu}\rangle.
\end{equation}
This equation is solvable if the nature of $\langle F^{\mu\nu}F_{\mu\nu}\rangle$ and $\langle\Delta^2\rangle$ are defined. Using the standard expressions for field quanta distributions
\begin{equation}\label{16}
\langle F^{\mu\nu}F_{\mu\nu}\rangle=-\dfrac{\nu}{2\pi^2}\int_0^\infty{dk\dfrac{k^4}{E_A}n_B(E_A)},
\end{equation}
and 
\begin{equation}\label{17}
\langle\Delta^2\rangle=\dfrac{1}{2\pi^2\phi^2_0}\int_0^\infty{dk\dfrac{k^2}{E_\chi}n_B(E_\chi) },
\end{equation}
with $n_B(x)=(e^{\beta x}-1)^{-1}$ the Bose-Einstein distribution function, $\beta=1/T$, where $T$ is the temperature and $\nu$ is the degeneracy of the gauge field. We solve these two equations by assuming relativistic limits where $E\approx k$ and $T\gg m$ \cite{Collins, Linde4,aIssifu} and thus the thermal fluctuation is written as:
\begin{align}\label{18}
\langle F^{\mu\nu}F_{\mu\nu}\rangle&=-\dfrac{\nu}{2\pi^2}\int_0^\infty{dk \dfrac{k^4}{E_A} \dfrac{1}{e^{\beta E_A}-1} }\nonumber \\
&=-\dfrac{\nu T^4}{2\pi^2}\int_0^{\infty}{\dfrac{x^3 dx}{e^x-1} }\nonumber\\
%&=-\dfrac{\nu  \rho^4T_g^4}{8\pi^2 m_A^4}\nonumber\\
&=-\dfrac{T^4}{T^4_{cg}}\langle F^{\mu\nu}F_{\mu\nu}\rangle_0\,,
\end{align}
where we have parameterized $x=k/T$ and also defined a critical temperature $T_{cg}$,
\begin{equation}\label{19}
T^4_{cg}=\dfrac{30 \langle F^{\mu\nu}F_{\mu\nu}\rangle_0}{\pi^2\nu}.
\end{equation}
Only the thermal fluctuations are taken into account in Eq.~\eqref{18} and therefore to have the full value of the gluon condensate it is necessary to add the ground state condensate, namely, $\langle F^{\mu\nu}F_{\mu\nu}\rangle_0$, and we get 
\begin{equation}\label{20}
\langle F^{\mu\nu}F_{\mu\nu}\rangle=\langle F^{\mu\nu}F_{\mu\nu}\rangle_0\left[ 1-\dfrac{T^4}{T^4_{cg}}\right].
\end{equation}
In effect, we have a maximum condensate at $T\,=\,0$, it decreases with increasing temperature and completely vanishes at $T=T_{cg}$. Also, Eq.~(\ref{17}) becomes 
\begin{align}\label{21}
\langle\Delta^2\rangle&=\dfrac{1}{2\pi^2\phi_0^2}\int_0^\infty{dk \dfrac{k^2}{E_\chi}\dfrac{1}{e^{\beta E_\chi}-1}}\nonumber\\
&=\dfrac{T^2}{2\pi^2\phi_0^2}\int_0^{\infty}{\dfrac{xdx}{e^x-1}}\nonumber\\
&=\dfrac{T^2}{3T_c^2},
\end{align}
where, $T_c^2=4\phi_0^2$, similar relation as in (\ref{21}) have been reached in \cite{Linde,Collins} using similar conditions. {We can also derive the Debye screening mass obtained from the leading term of the QCD coupling expansion, from the above expression, through
\begin{equation}\label{debye}
  m_D^2=m^2\langle\Delta^2\rangle=\dfrac{m^2T^2}{12\phi_0^2}=\tilde{g}^2T^2,
\end{equation}
with $\tilde{g}^2=m^2/12\phi_0^2$ a dimensionless coupling constant. We multiplied $\langle\Delta^2\rangle$ by $m^2$ to make it dimensionful. The $m_D=\tilde{g}T$ \cite{Kajantie, Manousakis} is the mass observed in the QGP region, showing its nonperturbative characteristics \cite{Shuryak, Niida}.} %This further reinforces the nonperturbative properties of the study.
Substituting Eqs.~(\ref{18}) and (\ref{21}) into Eq.~(\ref{14}), we obtain
\begin{equation}\label{17a}
    m^{*2}_\chi=m_A^2\left[1-\dfrac{T^4}{T^4_{cg}}\right]-m^2\left[1-\dfrac{T^2}{T^2_c}\right],
\end{equation}
here, $m_A^2=g^2\langle F^{\mu\nu}F_{\mu\nu}\rangle_0/\phi_0^2$ and $m^2=B_0/\phi_0^2$. 
In this case, the thermal fluctuating glueball mass vanishes at $T=T_{cg} \approx T_c$, at a point where the gluon condensate melts, %{\color{red} TF: Why $T_{cg}=T_c$?}, 
while the effective glueball mass obtained at $T=0$ corresponds to 
\begin{equation}\label{a3}
    m^{*2}_\chi(0)=m^2_A-m^2,
\end{equation}
for effective glueball mass at $T=0$, we expect $m_A\,>\,m$. 
%{\color{red} TF: what is the difference between the stable effective glueball mass at T=0 and the glueball mass?}
The gluon mass $m_A$ has been determined through lattice QCD simulation to be within the range $600\sim 700\,\text{MeV}$ \cite{Leinweber, Alexandrou, Langfeld} while other massive gluon masses %{\color{red} TF: are you referring to the gluon?} 
have also been determined through QCD phenomenology \cite{Field, Consoli} and QCD lattice studies \cite{Kogan} to be within $\sim 1\,\text{GeV}$. On the other hand, $m_A$ has been determined from Yang-Mills theory with an auxiliary scalar field $\phi$, to be $m_A=m(0^{++})/\sqrt{6}$ \cite{Cornwall, Kondo}, with $m(0^{++})$ the scalar glueball mass determined as fluctuations around $\phi$. Nonetheless, an alternative result was obtained in \cite{Issifu,aIssifu} to be $m_A=m_\phi/2$, with $m_\phi$ the scalar glueball mass. Based on the afore discussions, it is clear that in a system comprising $m$ and $m_A$, $m$ must always be greater than $m_A$. Thus, for $m_A=0$, we have a negative $m_\chi^{*2}$ representing the behavior of a glueball mass in the QGP region \cite{Susskind, Petreczky, Adams}. From QCD-based calculations, the $m_A$ obtained from non-vanishing gluon condensate is, 
\begin{equation}
    m_A^2=\left[\dfrac{34N\pi^2}{9(N^2-1)}\left<\dfrac{\alpha_s}{\pi}G^{\mu\nu}G_{\mu\nu}\right>\right]^{1/2},
\end{equation}
where $N$ is the color number, $\alpha_s$ is the strong coupling constant, and $G_{\mu\nu}$ is the non-abelian gauge field strength \cite{Gorbar, Felder:2001kt, Issifu}. In this study, we have approximated the non-abelian gauge field with an abelian one $F_{\mu\nu}$ coupled with $G(\phi)$, this approximation has been justified in \cite{aIssifu}. 
Since the critical temperatures ($T_{cg}\; \text{and}\; T_c$) are proportional to the corresponding masses ($m_A\; \text{and}\; m$) it follows that $T_{cg}\,>\,T_c$. Hence, we can determine the thermal fluctuating glueball potential by substituting Eq.~(\ref{17a}) into Eq.~(\ref{5d})
\begin{equation}\label{a2}
    V_c(T,r)=-\dfrac{q}{4\pi\varepsilon_0}\dfrac{\cot\left(\sqrt{m_A^2\left[1-\dfrac{T^4}{T^4_{cg}}\right]-m^2\left[1-\dfrac{T^2}{T^2_c}\right]}\,\,r\right)}{2\kappa^2\sqrt{ m_A^2\left[1-\dfrac{T^4}{T^4_{cg}}\right]-m^2\left[1-\dfrac{T^2}{T^2_c}\right]}}.
\end{equation}

 Again, Eq.~(\ref{17a}) has two critical temperatures at $T_{cg}$ and $T_c$, however, we can determine a common critical temperature $T_{c1}$ at which $m^*_\chi(T)$ vanishes $m^*_\chi(T=T_{c1})\,=\,0$ and determine $T_{c1}$, assuming $T_{cg}\,\approx\,T_c\,=\,1$. Then,
\begin{equation}\label{st2}
    {m_A^2\left[1-{T^4}\right]-m^2\left[1-{T^2}\right]}=0,
\end{equation}
solving the above equation for $T$, we get 
\begin{equation}\label{st3}
    T_{c1}=\sqrt{\dfrac{m^2-m_A^2}{m^2_A}}.
\end{equation}
With this newly defined $T_{c1}$ above, Eq.~(\ref{17a}) becomes
\begin{equation}\label{st5}
    m^{*2}_\chi(T)={m_A^2\left[1-\left(\dfrac{m^2_A}{m^2-m^2_A}\right)^2\dfrac{T^4}{T_{c1}^4}\right]-m^2\left[1-\left(\dfrac{m^2_A}{m^2-m^2_A}\right)\dfrac{T^2}{T_{c1}^2}\right]}.
\end{equation}

\begin{figure}[!t]
  \centering
 {\includegraphics[scale=0.5]{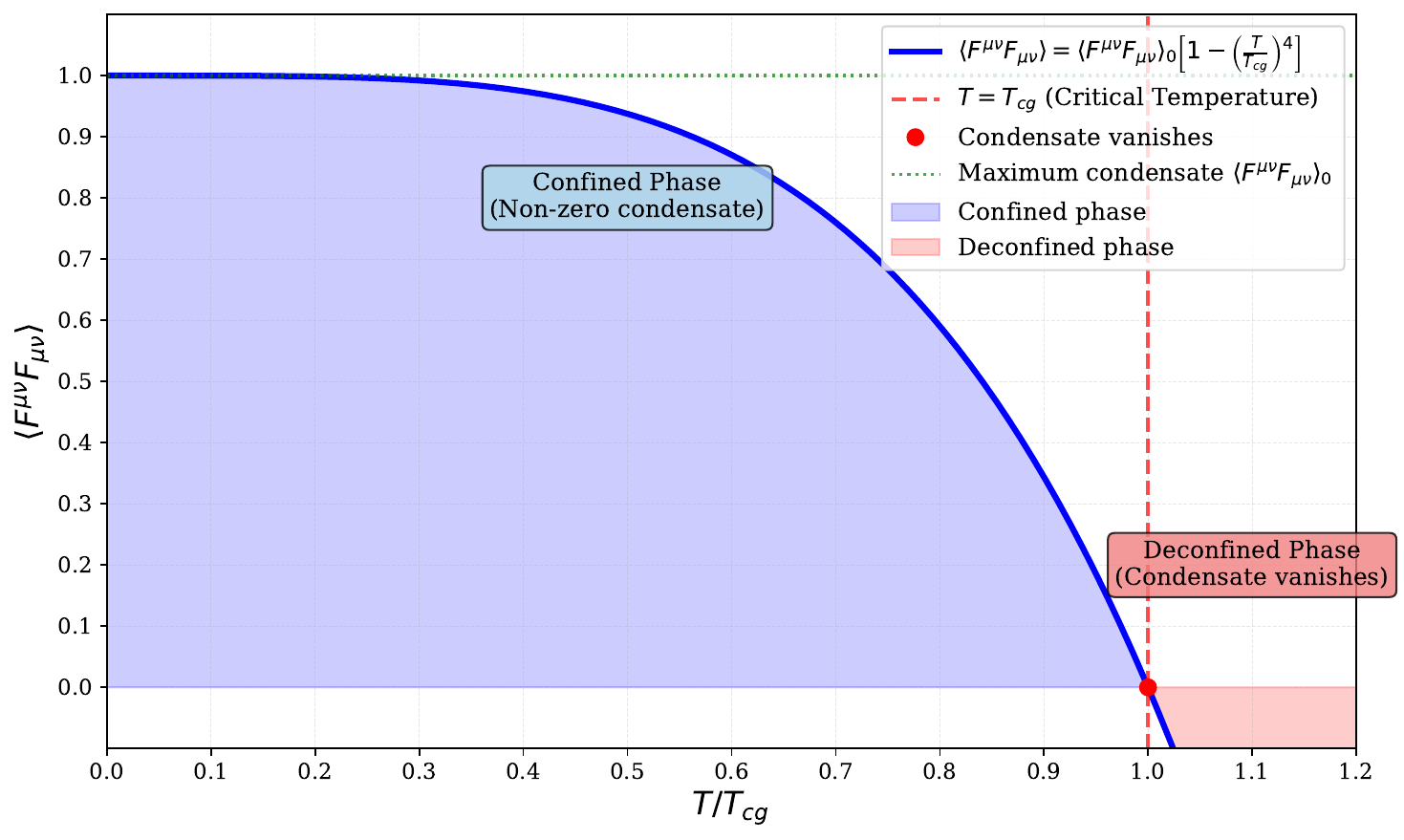}}
 \caption{A graph of $\langle F^{\mu\nu}F_{\mu\nu}\rangle/\langle F^{\mu\nu}F_{\mu\nu}\rangle_0$ against $T/T_{cg}$. Thus, the gluon condensate decreases rapidly with temperature and vanishes at $T=T_{cg}$.}
   \label{pc1}
\end{figure}

\begin{figure}[!t]
  \centering
 {\includegraphics[scale=0.5]{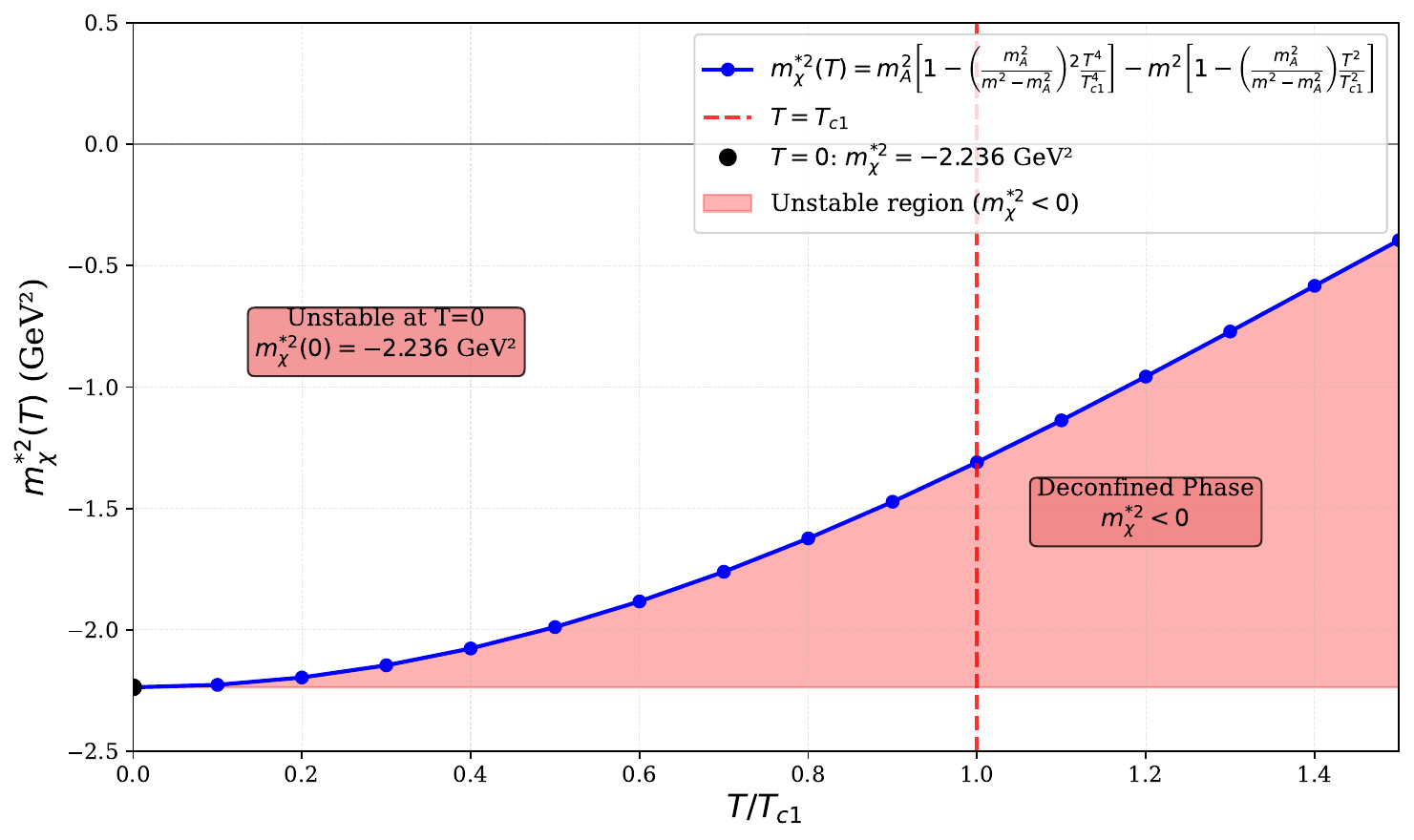}}
 \caption{The effective glueball mass, $m_\chi^{*2}(T)$, as a function of $T/T_{c1}$. The diagram is for $m\,=\,1.73\,\text{GeV}$ and $m_A\,=\,0.87\,\text{GeV}$, we find that the $m_\chi^*$ is unstable even at $T=0$, and beyond $T=T_{c1}$.}
   \label{pc2}
\end{figure}

\begin{figure}[!t]
  \centering
 {\includegraphics[scale=0.5]{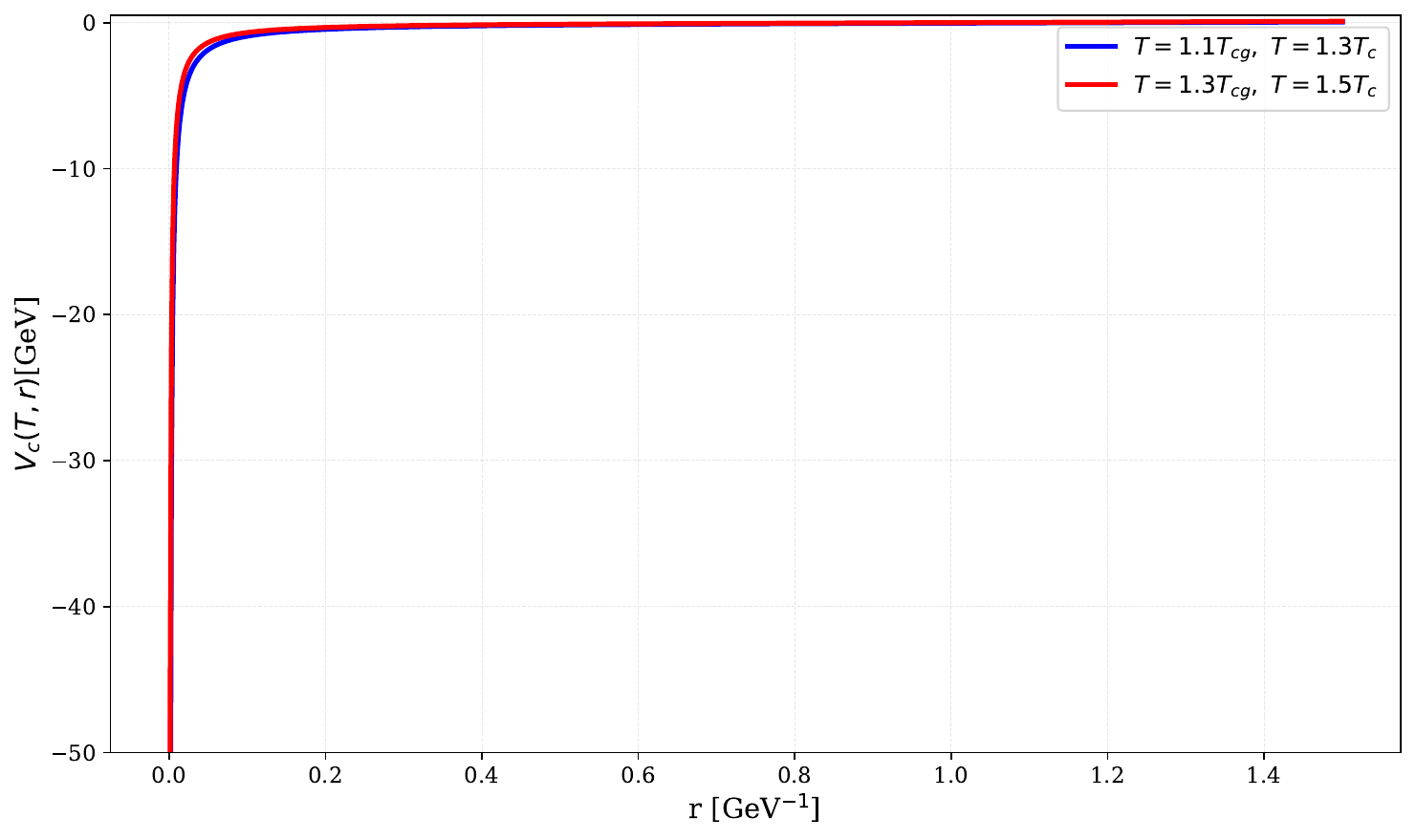}}
 \caption{ The effective potential, $V_c(r,T)$ (\cref{a2}) of the dense gluon-glueball admixed medium as a function of $r, T$. In this region, particles are already in a deconfined state with higher densities and temperatures. The temperature range here is beyond the critical temperature at which color confinement is expected. The graph is for $T_{cg}>T_c$. We observe that the potential lies in the asymptotically free region. Therefore, a change in temperature does not change the potential significantly.}
   \label{pc2a}
\end{figure}
The negative square mass (due to $m_A<m$ discussed above) implies the potential will have a negative curvature, which corresponds to the QGP region where the particles are in a deconfined state. This behavior remains unchanged even at $T=0$ because this is a high-density region where particles are asymptotically free. 

Figure~\ref{pc} shows the suppression of the gluon condensate $g^2\langle F_{\mu\nu}F^{\mu\nu}\rangle$ as a function of the dimensionless glueball mean field $\bar{\chi}$, obtained from Eq.~(\ref{12}). The condensate decreases monotonically with increasing $\bar{\chi}$, reflecting the back-reaction of the glueball field on the gluonic vacuum. As $\bar{\chi}\!\rightarrow\!1$, the condensate approaches zero, indicating the dissolution of the confining vacuum and the onset of deconfinement, consistent with Eq.~(\ref{20}), where the condensate vanishes at $T_{cg}$, displayed in Fig.~\ref{pc1}. The trend highlights the role of glueball fields as dynamical mediators between confined and deconfined phases, with direct implications for the thermodynamics of the early universe. Figure~\ref{pc2a} presents the confining potential $V_{c}(r,T)$ in the high-temperature, high-density regime where gluons are in a deconfined state. In this medium, the gluon condensate $\langle F^{\mu\nu}F_{\mu\nu}\rangle$ melts and the system transitions to a gluon-rich plasma. However, glueball-like excitations can still survive in this medium as collective, color-singlet excitations of the gluon field, even in the absence of permanent confinement. Thus in the model framework the glueball field $\phi$ remains dynamical in this regime with a negative square mass signaling instability towards symmetry restoration.  
The potential exhibits a flat, non-confining profile that remains largely insensitive to temperature variations in this phase. Signaling asymptotic freedom and the absence of confinement---behavior consistent with the QGP observed in heavy-ion experiments~\cite{Kajantie, Manousakis}.

\section{Cosmological Consequences of the model}\label{cosmology}
\subsection{Vacuum Energy Density}\label{cosmology1}
The stress tensor $T_{\mu\nu}$ of the glueball field $\phi$ is given by,
\begin{equation}\label{cp1}
T_{\mu\nu}=\partial_\mu\phi\partial_\nu\phi-\mathcal{L}_{eff}g_{\mu\nu}.
\end{equation}
Taking $\phi=\langle\phi\rangle$ to be constant, the equation reduces to $T^{\mu\nu}=V(\langle\phi\rangle)g^{\mu\nu}$, so the vacuum energy density reads 
\begin{equation}\label{cp2}
\langle T^0_0\rangle\equiv\rho_v=-\dfrac{m^4}{4\lambda}=-\dfrac{9\langle\alpha_s G_{\mu\nu}G^{\mu\nu}\rangle}{32\pi},
\end{equation}
we substitute \cref{csc} to obtain the term at the far right of the above expression. The contribution of vacuum energy density to the density of the universe is comparable to the critical density $\rho_c=2\times 10^{-29}h^2\text{kg}\text{m}^{-3}\simeq 10^{-46}\text{GeV}^4$ \cite{Linde5}. A large vacuum energy density, i.e., $\rho_v\gg\rho_c$, means the universe is decelerating rapidly such that it will contract again, so there will be no stars evolving. When $\rho_v\ll \rho_c $, the universe will expand rapidly, such that matter cannot condense into galaxies and stars.  In both instances, heavy elements would not have formed, and life would not have existed as we know it today. %current expansion rate is greater than the observational result. 
The result in \cref{cp2} has the wrong sign; such negative vacuum energy has never been measured and is nonphysical even though it is possible to some extent under quantum field theory. A universe with negative energy density is likely to violate energy conditions of general relativity and possibly the second law of thermodynamics \cite{Hawking, Nemiroff}. On the other hand, vacuum energy should be small relative to any fundamental energy scale; it can be $\rho_v=0$ but not negative. Thus, the challenge is solved when we use the $V_{eff}(\phi, T)$ for $T>T_{c\phi}$ i.e.
\begin{equation}\label{cp4}
\langle T^0_0\rangle\equiv\rho_v=\dfrac{9\langle\alpha_s G_{\mu\nu}G^{\mu\nu}\rangle}{32\pi}\left[\dfrac{T^4}{T^4_{c\phi}}-1 \right].
\end{equation}
%{\color{red}TF: is this equation related to Eq. 52? Explain}
{The term in the square bracket comes from the thermal correction derived from integrating out the gauge field in \cref{1} and $\langle \alpha_sG_{\mu\nu}G^{\mu\nu}\rangle$ comes from substituting $\lambda$ with \cref{csc}. This expression is analogous to \cref{20}.}
At $T=0$ we recover \cref{cp2} with the wrong sign, when we set $T=T_{c\phi}$ we have $\rho_v=0$, the minimum possible vacuum energy density, and at $T>T_{c\phi}$ we have the correct sign for $\rho_v$.

\subsection{Domain Wall}\label{cosmology2}
As discussed extensively above, at sufficiently high temperatures $T\geq T_{c\phi}$ the symmetry of the $V_{eff}(\phi,T)$ is restored. With a decrease in temperature below $T_{c\phi}$, SSB reappears. Here, we suppose that the space is divided into two regions: one in the state $\phi_0=-m/\sqrt{\lambda}$ and the other in the state $\phi_0=m/\sqrt{\lambda}$, with respect to the degenerate vacuum. Since we expect $\phi$ to make a smooth transition from  $\phi=-m/\sqrt{\lambda}$ to $\phi=m/\sqrt{\lambda}$, it must go through the false vacuum state $\phi_0=0$. This transition is referred to as the domain wall \cite{Kolb, Linde5}. These domains are separated from each other by thin walls in $\phi_0=0$, where the field changes from  $\phi_0=-m/\sqrt{\lambda}$ to  $\phi_0=m/\sqrt{\lambda}$. These walls have a sufficiently large surface energy density, so if any of these walls existed presently in the observable part of the universe, it would make the universe largely anisotropic (this seems obvious in standard scenarios). That means potentials with SSB, such as \cref{2}, do not conform with cosmological data without any corrections introduced. That notwithstanding, theories with SSB such as the  \textit{Grand Unified theory} ($\text{SU}(5)$) \cite{Collins} theories with spontaneous broken {\it CP invariance}, {\it Weinberg model with  CP violation} \cite{Weinberg76}, some \textit{axion theories} \cite{Axiom, Abbott:1982af} and many more, remain important in elementary particle physics today. Most of these theories are suitable in many areas of physics, so an effort to save at least some of them is a novelty worth exploring. %{\color{blue}
Thus, the \( V_{\text{eff}}(\phi, T) \) with the thermal correction to \( \phi^2 \) allows us to resolve the domain-wall problem for \( T > T_{c\phi} \) in \cref{39}, due to the sign change in \cref{38} within this temperature range, yielding
\begin{equation}\label{39b}
m^{2}(T) = -m^{2}\left[\frac{T^{4}}{T_{c\phi}^{4}} - 1 \right].
\end{equation}
In this scenario, the effective potential develops a global minimum at \( \phi_{0} = 0 \), thereby suppressing the formation of domain walls. This minimum persists until \( T = T_{c\phi} \), keeping the Universe sufficiently hot and free of domain walls as the phase transition proceeds.
Many different approaches have been proposed for resolving problems of this kind. Common among which is the introduction of thermal distribution to $\phi$ such that $\phi\rightarrow \phi+\phi_T$. This modifies the Higgs-like potential in \cref{2} such that it acquires minima $\phi_0 =0$ \cite{Collins,Linde4,aIssifu,Issifu} at $T\geq T_{c\phi}$.  The main difference in our approach compared to others is that we introduced radiation correction from the gauge field into the scalar potential instead of the scalar field to the scalar field approach. With the gauge field, the temperature correction is in the order of $\sim T^4$ while the scalar field is in the order $\sim T^2$ (the two forms of temperature introduction have been discussed in \cite{aIssifu}). 

\subsection{The Inflationary Universe}\label{cosmology3}
The most general form of Einstein's equations of motion, including the cosmological constant, is given as 
\begin{equation}\label{iu1}
R_{\mu\nu}-\dfrac{1}{2}g_{\mu\nu}\mathcal{R}=8\pi G_NT_{\mu\nu}+\Lambda g_{\mu\nu},
\end{equation}
where $R_{\mu\nu}$ is the Ricci tensor, $\mathcal{R}$ is the Ricci scalar, $T_{\mu\nu}$ is the energy-momentum tensor, $G_N$ is the Newton's gravitational constant and $\Lambda$ is the cosmological constant. The cosmological constant acts as an additional form of stress-energy with a constant vacuum energy density and isotropic pressure at $\phi_0=\pm m/\sqrt{\lambda}$, thus 
\begin{equation}\label{iu2}
\rho_v=\frac{\Lambda}{8\pi G_N} \qquad{\text{and}}\qquad p_v=-\dfrac{\Lambda}{8\pi G_N},
\end{equation} 
respectively. Using \cref{cp4} at $T>T_{c\phi}$, we obtain 
\begin{equation}\label{iu3}
\Lambda=\dfrac{2\pi G_N m^4}{\lambda}\left[\dfrac{T^4}{T^4_{c\phi}}-1 \right].
\end{equation}
We used the relations $\rho=(1/2) \dot{\phi}^2+V_{eff}(\phi,T)$ and $p=(1/2)\dot{\phi}^2-V_{eff}(\phi,T)$, where $\rho_v$ and $p_v$ were calculated in the vacuum $\phi_0$ with the equation of state $\omega=p/\rho=-1$. At $T=T_{c\phi}$, $\Lambda=0$, we have an expanding universe as presented by Alexander Friedmann \cite{Friedman}, later affirmed by Hubble's observation \cite{Hubble}, and finally accepted by Einstein \cite{Einstein}. At $T>T_{c\phi}$ we have $\Lambda>0$, consistent with the modern observational cosmology which points to a positive cosmological constant \cite{observation, Boomerang:2000efg, Hanany:2000qf, Boomerang:2000jdg} and confirmed by satellite observation \cite{Spergel}. At $T=0$, $\Lambda<0$, but it is important to point out that at this temperature regime, there is an extremum at $\phi_0=0$ and a global minimum at $\phi_0=\pm m/\sqrt{\lambda}$, so the true value of $\Lambda$ must take into consideration the maxima. Thus, $\rho_v=V(0)-V(\langle\phi\rangle)$ making $\Lambda>0$, this holds for $0\leq T<T_{c\phi}$. Consequently, before the phase transition $T=T_{c\phi}$, there is super-cooling at $T=0$ with an effective cosmological constant 
\begin{equation}\label{iu4}
\Lambda=\dfrac{2\pi G_N m^4}{\lambda}.
\end{equation}
Now, substituting Eq.~(\ref{csc}) into the above relation yields,
\begin{equation}\label{q1}
    \Lambda=\dfrac{9G_N\left\langle\alpha_sG_{\mu\nu}G^{\mu\nu}\right\rangle}{4}.
\end{equation}
Hence, the cosmological constant is related to the gluon condensate and has its maximum value at $T=0$. In this region, \textit{color confinement} of the glueballs is observed, similar to the discussions in Sec.~\ref{CG}. 

\begin{figure}[!t]
    \centering
    \includegraphics[scale=0.5]{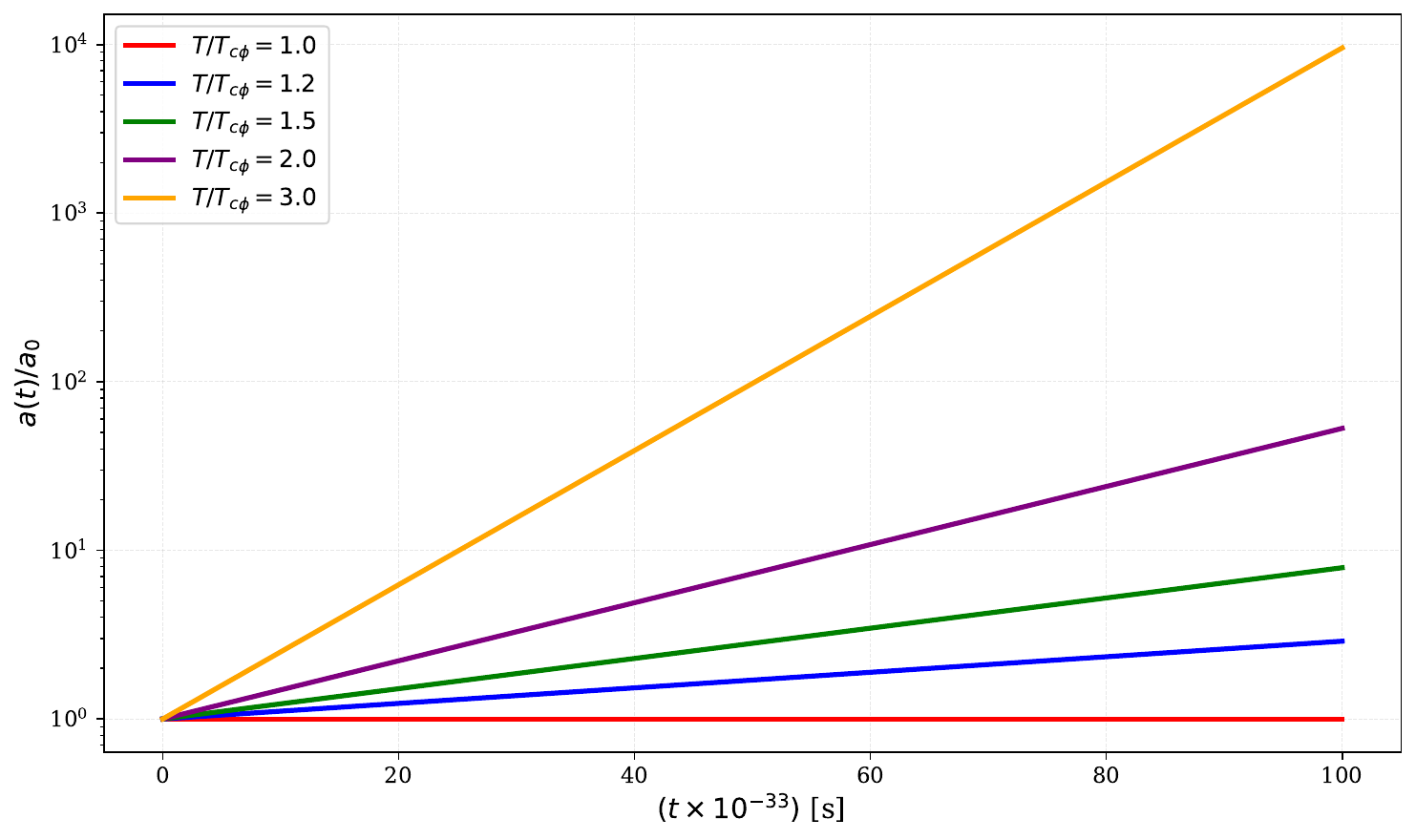} %
     \caption{A diagram of \( a(t)/a_0 \) is shown for different values of \( t \) (in seconds) during the inflationary epoch. It illustrates the evolution of the scale factor in a universe dominated by a cosmological constant and populated with deconfined glueballs. The gluon condensate significantly affects the dynamics of \( a(t) \), and it grows steadily with time.  
}%
     \label{fig:at}
 \end{figure}

The history of the $\Lambda$ has long-standing implications starting from its introduction by Einstein to develop a consistent model of a static universe to the modern-day discovery of acceleration in cosmic expansion, technical measurements of cosmic geometry, recent advances in galaxy formation and galaxy clustering, etc. \cite{Baez, Raifeartaigh}.
Substituting the expression for $\rho_v$ into the Friedman equation,
\begin{eqnarray}
H^{2}=\dfrac{8\pi G_N}{3}\rho_v +\dfrac{\Lambda}{3},
\end{eqnarray}
substituting (\ref{iu2}) gives
\begin{eqnarray}\label{ju}
H^{2}_\Lambda=\dfrac{2\Lambda}{3},
\end{eqnarray}
%{\color{red} TF: My understanding is that you don't want to include and adhoc cosmological $\Lambda/3$ in the Friedmann equation, then Eq. (71) would have only the term $H^2=\dfrac{8\pi G_N}{3}\rho_v$, and the previous discussion was to obtain $T_{00}$ or $\rho_v$ from your Lagrangian model Eq. (1) {\color{blue} Eq.~(\ref{35})} including the thermal fluctuations and vacuum value of $\langle F_{\mu\nu}F^{\mu\nu} \rangle$ reinterpreted as the expectation value of the gluon condensate $\langle G_{\mu\nu}G^{\mu\nu} \rangle$. So Eq.72 would be $H^{2}_\Lambda=\dfrac{\Lambda}{3}$ derived from the $8\pi G_N \rho_v/3$. It is correct my understanding? } {\color{blue}Correct! %We substituted Eq.~(67) for $\rho_v$ in (71) which is the complete Friedman equation with a cosmological constant to obtain (72).
%}
where $H=H_\Lambda=\dot{a}/a$ is the Hubble parameter and dot over $a$ is the derivative with respect to time. Solving for $a$ gives 
\begin{eqnarray}
a(t) = \exp\left({H_{\Lambda}}t\right) = \exp\left(\sqrt{\frac{2{\Lambda}}{3}}t\right),
\end{eqnarray}
where the integration constant $a(t_{0})=1$ is the scale factor of the early universe and $t_0\sim 0$ is some time in the early universe. %{\color{green} Great, thanks !} 
{Now substituting Eqs.~(\ref{csc}) and (\ref{iu3}) into the above expression yields the following
\begin{equation}\label{ifi}
    a(t)= \exp\left(\sqrt{\dfrac{3G_N\left\langle\alpha_sG_{\mu\nu}G^{\mu\nu}\right\rangle}{2}\left[\dfrac{T^4}{T_{c\phi}^4}-1\right]}t\right).
\end{equation}
Hence, we have an inflationary expanding universe when $T\,>\, T_{c\phi}$, with a nonzero gluon condensate corresponding to a unique vacuum of $V_{eff}(\phi, T)$. At $T =T_{c\phi}$, inflation stops with a disappearance of the gluon condensate corresponding to a unique vacuum of $V_{eff}(\phi, T)$. At $T\,<\,T_{c\phi}$, where $V_{eff}(\phi, T)$ has a degenerate vacuum, the universe oscillates through various cycles of collapse and expansion \cite{Dabrowski, Graham}.} 
The evolution of the scale factor $a(t)/a_{0}$ in Fig.~\ref{fig:at} reflects the link between glueball confinement dynamics and cosmology in our model. Different slopes correspond to inflationary phases driven by the gluon condensate, which acts as an effective cosmological constant $\Lambda$. For $T > T_{c\phi}$, a large $\Lambda$ (see Eq.~(\ref{iu3})) produces exponential expansion, while the transition to $T \leq T_{c\phi}$ ends inflation as $\Lambda$ vanishes and glueballs enter the confined phase, directly connecting the QCD deconfinement transition to early-universe inflation.

It is important to emphasize that the apparent differences between the QCD glueball mass scale ($\sim 1$--$2~\mathrm{GeV}$) \cite{Cheng:2015iaa} and the much higher inflationary scale ($\sim 10^{16}~\mathrm{GeV}$ in GUT-scale models) \cite{Linde} is naturally resolved within our model framework. In our scenario, the inflaton field $\phi$ does not derive its inflationary energy from the glueball mass of the confined phase, but from the much larger vacuum energy of the gluon condensate in the high-temperature deconfined regime. Specifically, the inflationary energy scale is controlled by the thermal correction term proportional to $\langle \alpha_s G_{\mu\nu}G^{\mu\nu}\rangle$ and $T^4$ in Eq.~(\ref{ifi}), such that $\phi$ drives inflation through the thermal gluon-plasma potential $V_{eff}(0,T)$ for $T > T_{c\phi}$ (as illustrated in Fig.~\ref{sp}, where the potential exhibits a unique minimum at $\langle\phi\rangle=0$). The field only acquires its low-energy glueball mass after the phase transition at $T < T_{c\phi}$, when the symmetry breaks spontaneously and the field settles into the confined-phase minimum of the now degenerate vacuum of $V_{eff}(\langle\phi\rangle,T)$.

\subsection{Fitting the Model into the Planck Data}\label{cosmology4}

  From the effective potential in Eq.(\ref{36}), we can conveniently write
\begin{eqnarray}
V_{eff}(\phi)=\dfrac{\lambda}{4}\phi^{4}-\dfrac{m^2}{2}\phi^{2}+\dfrac{\sigma^{2}}{3}\phi^{2},
\end{eqnarray}
where 
\begin{equation}\label{cm2}
\sigma^2=\dfrac{4\kappa^2\pi^2T^4}{15},
\end{equation}
is the thermal correction calculated from the gauge field. We can calculate the spectral index of the inflaton $\phi$ from the slow-roll parameters \cite{Lyth}. 
\begin{equation}\label{cm3}
\epsilon=\dfrac{M_{p}^{2}}{2}\left( \dfrac{V_{eff,\phi}}{V_{eff}}\right)^{2}, \qquad\mbox{and}\qquad \eta=M_{p}^{2}\dfrac{V_{eff,\phi\phi}}{V_{eff}}.
\end{equation}
{where $V_{eff,\phi}$ represents the first derivative of the effective potential  and $V_{eff,\phi\phi}$ is the  second derivative.}

The results are
\begin{eqnarray}
\epsilon &=& \frac{8M_{p}^{2}}{\phi^{2}}\left(\frac{3\lambda\phi^{2}-\rho^{2}}{3\lambda\phi^{2}-2\rho^{2}}\right)^{2}\\
\eta &=& \frac{4M_{p}^{2}}{\phi^{2}}\left(\frac{9\lambda\phi^{2}-\rho^{2}}{3\lambda\phi^{2}-2\rho^{2}}\right)\,\mbox{,}
\end{eqnarray}
where we define $\rho^{2} = 2\sigma^{2}-3m^{2}$. Now, considering $\tau = \lambda/\rho^{2}$, and expanding in first order in $\tau$, we get

\begin{eqnarray}
\epsilon &=& \frac{2M_{p}^{2}}{\phi^{2}} - 6\gamma\\
\eta &=& \frac{2M_{p}^{2}}{\phi^{2}} - 15\gamma\,\mbox{,}
\end{eqnarray}
with $\gamma = \tau M_{p}^{2}$ and $M_p$ the Planck mass. The spectral index $n_s$ and tensor to scalar ratio $r_s$  can also be calculated from the expressions 
\begin{equation}\label{cm4}
n_s=1-6\epsilon+2\eta \quad{,}\quad r_s=16\epsilon,
\end{equation}
consequently,
\begin{eqnarray}\label{ns}
n_{s} = 1-\frac{8M_{p}^{2}}{\phi^{2}} + 6\gamma,
\end{eqnarray}
\begin{eqnarray}\label{r}
r_{s} = \frac{32M_{p}^{2}}{\phi^{2}} - 96\gamma.
\end{eqnarray}

Furthermore, the e-folds number, $N$ is given by
\begin{eqnarray}
\label{Neff}
N = \frac{1}{M_{p}^{2}}\int_{\phi_{e}}^{\phi_{k}}\frac{V_{eff}}{V_{eff,\phi}}\,d\phi,
\end{eqnarray}
where $\phi_{e}$ represents the scalar field value at the end of inflation. $\phi_k$ is the value of the scalar field at horizon crossing for a given Fourier mode $k$, defined by the condition $k = aH$, and it determines the initial conditions for the corresponding curvature perturbations. This value is achieved by the condition $\epsilon(\phi_{e}) = 1$, what implies that
\begin{eqnarray}
\label{phi_e}
\phi_{e}^{2} = \frac{2M_{p}^{2}}{1+6\gamma}.
\end{eqnarray}
Therefore, from Eqs. (\ref{Neff}) and (\ref{phi_e}), we obtain
\begin{eqnarray}
\label{Ngamma}
N = \frac{\phi_{k}^{2}}{4M_{p}^{2}}+\frac{3\gamma}{16}\frac{\phi_{k}^{4}}{M_{p}^{4}} - N_{e}\,\mbox{,}
\end{eqnarray}
with \[N_{e} = \frac{2+15\gamma}{4(1+6\gamma)^{2}}\,\mbox{.}\]
Rewriting $\phi_{k}$ as a function of $N$ and $\gamma$, we can write the $n_s$ and $r_s$ as follows
\begin{eqnarray}
\label{nsrplane}
n_s &=& 1+6\gamma + \frac{12\gamma}{1-\sqrt{1+12\gamma(N+N_{e})}}, \\
r_s &=& \frac{48\gamma}{-1+\sqrt{1+12\gamma(N+N_{e})}} - 96\gamma.\label{nsrplane1}
\end{eqnarray}

\begin{figure}[!t]
    \centering
    \includegraphics[scale=0.8]{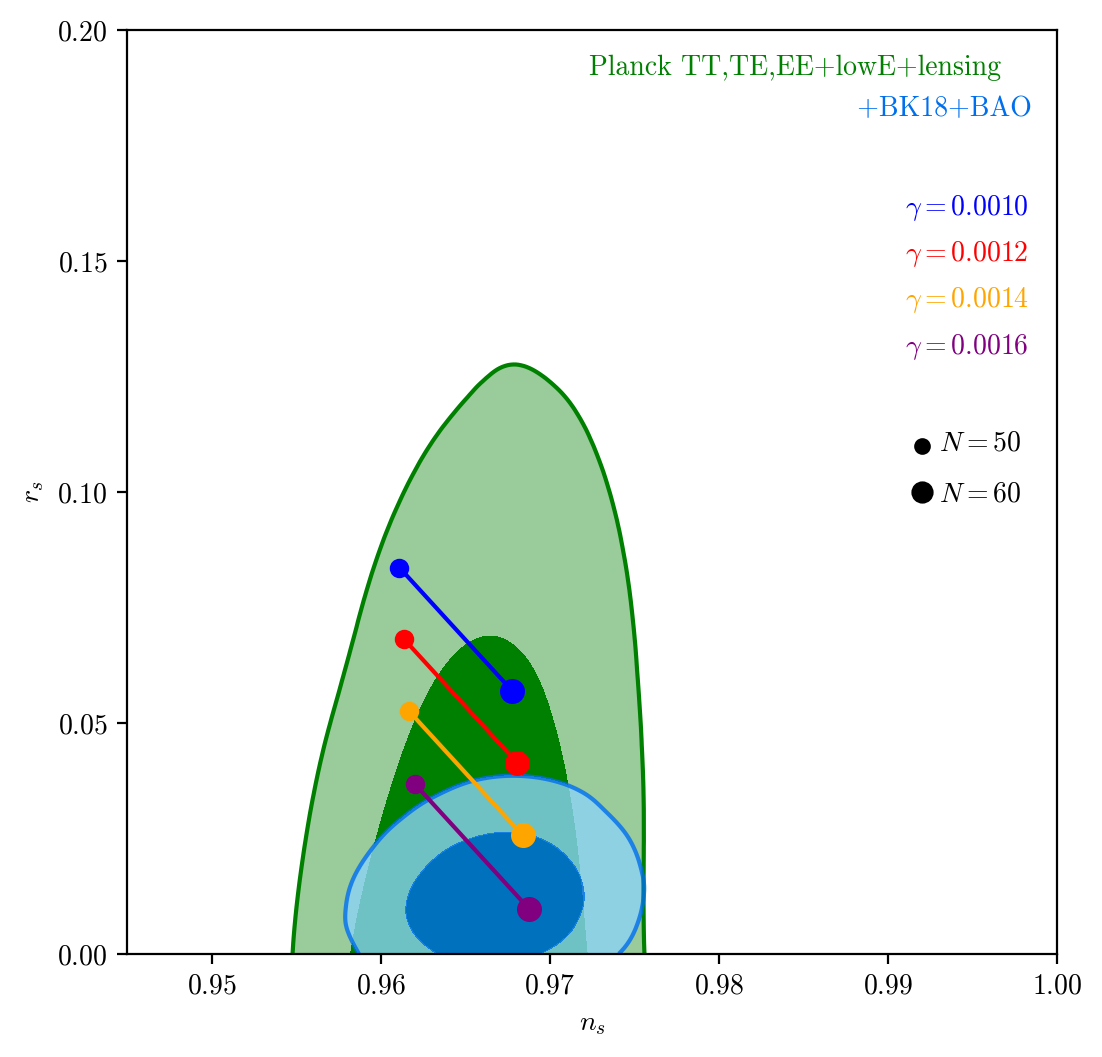}
    \caption{Predicted spectral index $n_s$ and tensor-to-scalar ratio $r_s$ in the glueball-inflation model, compared with Planck~2018 constraints (68\% and 95\% C.L.). The colored curves show predictions for different values of the coupling $\gamma$ and e-folds $N \simeq 50\text{--}60$, exhibiting the characteristic linear relation $r_s = 4(1 - n_s) - 72\gamma$. The agreement with Planck data indicates that glueball-driven QCD-scale dynamics can support a viable inflationary epoch.
}
    \label{fig:example}%
\end{figure}

On the other hand, the combined expressions of Eqs.~(\ref{nsrplane}) and (\ref{nsrplane1}) leads a linear fit linking the spectral indices to the  coupling $\gamma$:
\begin{eqnarray}\label{fit1}
r_s = 4(1-n_s)-72\gamma.
\end{eqnarray}
For Planck data \cite{Planck1,PhysRevLett.127.151301}, $n_s = 0.9658$ and $r_s<0.036$, one constrains $\gamma > 0.0014$. Moreover, the plane $n_s - r$ is shown in the Figure \ref{fig:example}. We can also express the power spectrum as
\begin{eqnarray}\label{pow}
P_{s}=\dfrac{V_{eff}}{24\pi^{2}M_{p}^{4}\epsilon},
\end{eqnarray}
such that if we consider the pivot scale to be $k=0.05\,\mbox{Mpc}^{-1}$, then $P_{s}=A_{s}=2.09\times 10^{-9}$ as obtained from Planck data. Hence, Eq.(\ref{pow}) becomes
\begin{eqnarray}\label{cm1}
A_{s} &=& \frac{\phi_{k}^{4}\rho^{2}}{2304\pi^{2}M_{p}^{6}}\frac{(3\tau\phi_{k}^{2}-2)^{3}}{(3\tau\phi_{k}^{2}-1)^{2}}\nonumber\\
&\approx & -\frac{1}{576\pi^{2}}\left(\frac{\phi_{k}}{M_{p}}\right)^{4}\left(\frac{\rho}{M_{p}}\right)^{2}\left(2+3\gamma\frac{\phi_{k}^{2}}{M_{p}^{2}}\right)
\end{eqnarray}

Considering $N=55.7$ and $\gamma = 0.0014$, the expression (\ref{cm1}) leads to result
\begin{eqnarray}
    \rho^{2} &=& 2\sigma^{2} - 3m^{2} = -0.0057M_{p}^{2}\\
    \lambda &=& \frac{\gamma\rho^{2}}{M_{p}^{2}} = - 7.96\times 10^{-5}
\end{eqnarray}

To confront our model with observational data, we fit the inflationary predictions derived from the $V_{eff}(\phi, T)$ to the latest Planck satellite constraints. By computing the slow-roll parameters and expressing the spectral index $n_s$, the tensor-to-scalar ratio $r_s$, and the running of the spectral index $\alpha_s$ in terms of the number of e-folds $N$, we establish a direct link between the glueball-inspired inflaton dynamics and cosmological observables. As shown in Fig.~\ref{fig:example}, the model’s predictions in the $n_s\!-\!r_s$ plane align with the Planck confidence regions for a range of values of the parameter $\gamma$ and for $N \approx 50\text{--}60$, demonstrating the viability of the framework. This agreement indicates that QCD confinement dynamics, encoded in the scalar glueball field, can naturally support a successful inflationary phase.

This consistency is complemented by the behavior of the scale factor $a(t)/a_0$ in Fig.~\ref{fig:at}, where the exponential expansion observed in the regime $T > T_{c\phi}$ signifies the inflationary phase driven by the gluon condensate. Together, these results show that the deconfinement transition of glueballs not only triggers the correct inflationary dynamics but also yields predictions for cosmological observables that agree with precision CMB measurements. The linear relation $r_s = 4(1 - n_s) - 72\gamma$, characteristic of the model, provides an additional testable signature for future observations. Furthermore, normalizing the scalar power spectrum to the observed value $A_s = 2.09 \times 10^{-9}$ allows us to constrain both the thermal correction parameter $\sigma^2$ and the dimensionless coupling $\gamma$. This analysis shows that the model not only reproduces the correct order of magnitude for the inflationary observables but also provides a viable framework for constraining nonperturbative QCD parameters through precision cosmology.

The parameter $\gamma$, as defined in Eq.~(\ref{fit1}), serves as a dimensionless coupling that links the QCD-inspired potential to the thermal gauge sector. $\gamma$ is connected to the thermal correction parameter $\sigma$ as $\gamma\propto 1/\sigma^2$, and since $\sigma^2\propto T^4$, a higher $T$ implies a smaller $\gamma$ and vice-versa. Therefore, a constrained small value, $\gamma > 0.0014$, provides an important consistency check of the model’s premise (discussed in the last paragraph of Sec.~\ref{cosmology3}): that the inflationary scale is sourced by the high-temperature gluon condensate rather than the low-energy glueball mass. A low $\gamma$ ensures a weak, natural coupling that keeps the predicted $(n_s, r_s)$ within Planck bounds without fine-tuning. In contrast, a large $\gamma$ would imply an unrealistically strong coupling (due to low $T$), distort slow-roll dynamics, and violate CMB constraints. Thus, the small allowed range of $\gamma$ supports the viability of the gluon condensate as the driver of both confinement dynamics and a successful inflationary epoch.

In the present framework, an inflationary epoch occurs only for $T > T_{c\phi}$, since in this temperature regime thermal corrections restore the symmetry of the effective potential, yielding a single stable vacuum at $\langle \phi \rangle = 0$ with a large, positive effective cosmological constant $\Lambda$. This vacuum energy dominates the cosmic expansion and drives an exponential growth of the scale factor, $a(t) \propto e^{H_{\Lambda} t}$, thereby satisfying the conditions for inflation. For $T \leq T_{c\phi}$, the system transitions to the confined phase: the cosmological constant vanishes, the potential develops degenerate minima, and rapid field oscillations, together with domain formation, preclude any sustained period of accelerated expansion. Inflation is therefore intrinsically linked to the high-temperature, deconfined phase of QCD.

%{\color{red} TF: To achieve the inflationary universe for $T>T_{c\phi}$ it is necessary to assume the analytic extension of Eq. 64 beyond the T that the gluon condensate vanishes, which is the key point of this work. This is still a surprise for me. Is it possible to provide a physical argument that this extension is possible?}

\subsection{(De)confinement and Cosmological Phase Transitions}\label{confinement}

In this subsection, we examine the behavior of glueballs across different temperature regimes and the associated cosmological phase transitions. We aim to identify the stages in the evolution of the universe during which glueballs existed in either a confined or deconfined state as determined by temperature variations. In Sec.~\ref{CG}, we showed that confinement is governed by the glueball mass $m$ appearing in $V(\phi)$. In Sec.~\ref{thermal}, we derived the effective thermal potential $V_{eff}(\phi, T)$, which forms the basis for analyzing high-temperature cosmological effects. By substituting the temperature-dependent 
glueball mass $m(T)$ obtained in Eq.~(\ref{38}) into Eq.~(\ref{5d}), we now obtain the \textit{thermal fluctuating glueball confining potential}:

\begin{align}\label{pr5}
V_c(r,T)=-\dfrac{q}{4\pi\varepsilon_0}\dfrac{\cot(m(T)r)}{2m(T)\kappa^2}.
\end{align}
The glueball mass vanishes at $T = T_{c\phi}$, decreases for $T < T_{c\phi}$, and at $T = 0$ we recover the confining potential given in Eq.~(\ref{5d}). From the above expression, we observe that $T = 0$, corresponding to $\Lambda > 0$, represents \textit{color confinement} of glueballs --- see Fig.~\ref{pba}. Likewise, $T = T_{c\phi}$, where $\Lambda = 0$, marks the phase-transition point at which $V_c(r, T)$ becomes nonphysical. This corresponds to the stage in the inflationary 
universe where ``super-cooling'' occurs due to tunneling, as discussed below Eq.~(\ref{36}). In terms of $V_c(r, T)$, this point signifies the transition of glueballs from a confined to a deconfined state.

\begin{figure}[!t]
  \centering
 \includegraphics[scale=0.35]{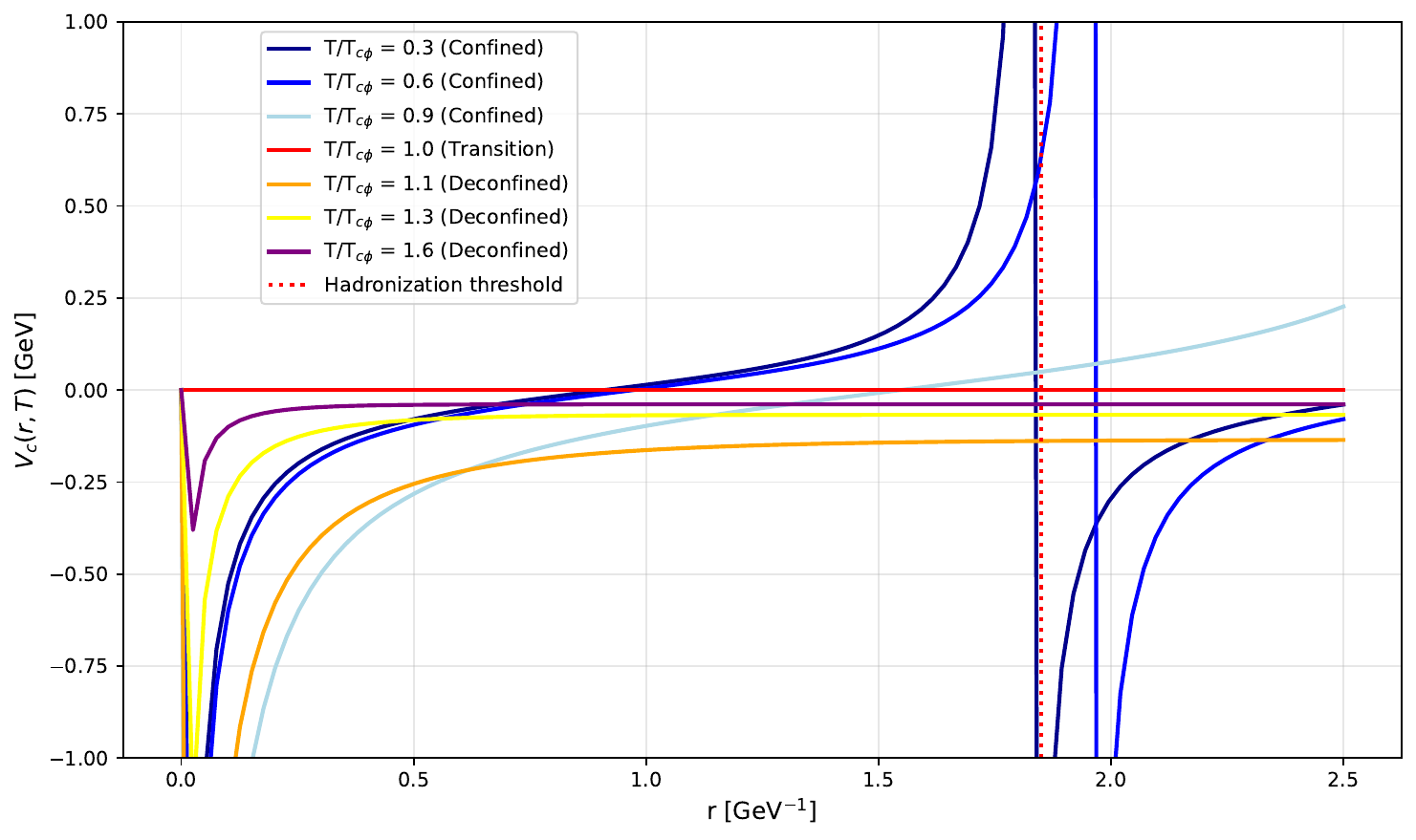}
 \includegraphics[scale=0.35]{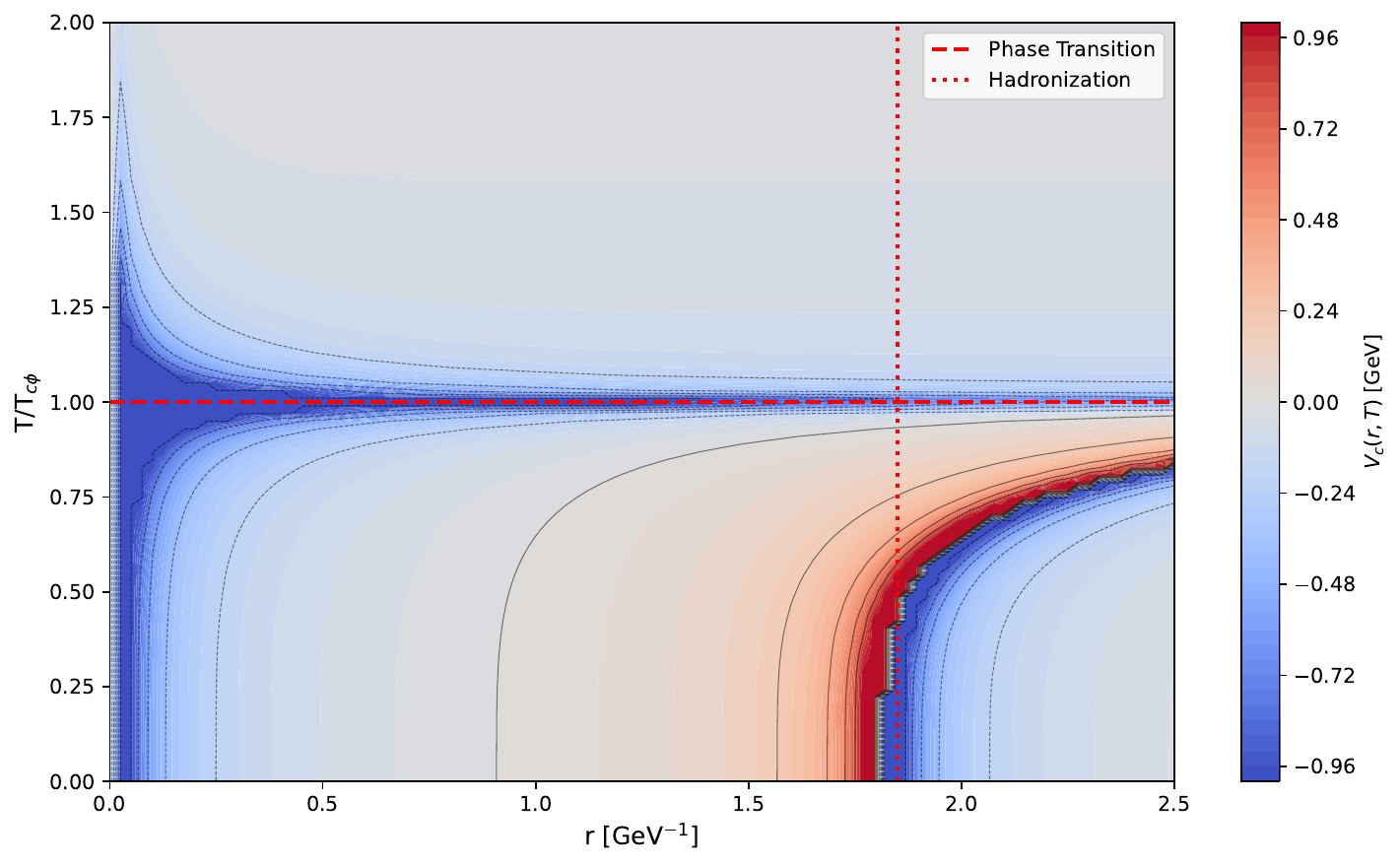}
 \caption{A graph of $V_c(r,T)$ against $r,T$ for fix values of $T/T_{c\phi}$ and varying $T/T_{c\phi}$. The potential shows a linear increase with decreasing temperature $T$ signaling \textit{color confinement}. At $T=T_{c\phi}$, there is a phase transition from a confined glueball phase to a deconfined glueball phase. Likewise, the gradient of the graph decreases with increasing $T$, attaining a negative gradient at $T\,>\, T_{c\phi}$ showing phase transition to a deconfined matter state. The vertical red dotted line in the graph represents the confining threshold beyond which hadronization occurs.}
   \label{pd}
\end{figure}

For $T > T_{c\phi}$, corresponding to $\Lambda > 0$, the system enters a fully deconfined phase. At these temperatures, the potential exhibits a characteristic phase-transition behavior with a negative slope, as shown in Fig.~\ref{pd}. This behavior is consistent with modern observational cosmology, established features of the early universe, and results from relativistic heavy-ion collision experiments at RHIC and the LHC~\cite{Shuryak, Niida}.

 At the transition temperature $T_{c\phi}$, the glueball mass that drives the confinement and deconfinement transitions vanishes. At this point, there are no particles in the system, so the potential shows nonphysical behavior. Just above, $T_{c\phi}$, the glueball mass becomes unstable and acquires a negative square mass. However, it has been established that at the quark-gluon-plasma (QGP) region, where the particles are in a deconfined state, both the gluon condensate and the isoscalar glueball mass change sign \cite{Kochelev2, Mathieu}. QCD lattice calculation of $\text{SU}(3)$ shows the same results \cite{Miller, Min:2007hc}. Since the glueball mass and the gluon condensate are related in Eq.(\ref{csc}), it implies that a change in the sign of the gluon condensate leads to a corresponding change in the sign of the glueball mass.

\section{Analysis and Conclusion}\label{analysis}

\subsection{Analysis}
In elementary particle theory, gauge fields mediate interactions between the constituent particles. In the present framework, the scalar glueball field $\phi$ representing color-singlet gluonic bound states, interacts through an effective Abelian gauge sector coupled via the color‑dielectric function $G(\phi)$. Generally, strong interactions are mediated by non‑Abelian gauge theory; the Abelian approximation is motivated by the established phenomenon of Abelian dominance in the confining regime of QCD, where the long‑distance dynamics are largely captured by Abelian degrees of freedom. The function $G(\phi)$ encodes the nonperturbative infrared behavior of the gluon field by effectively screening the chromoelectric flux at large separations and preventing the decoupling of the gauge sector. This minimal setup enables us to capture the self-interacting nature of gluons in the infrared while retaining analytical control over the confinement dynamics and phase-transition properties of the glueball system. The free parameters that require fixing in the model framework are; the glueball decay constant $\kappa=0.5\,\text{GeV}^{-1}$, the electric charge $q=+1$, vacuum permittivity $\varepsilon_0=1$, glueball mass $m=1.73\,\text{GeV}$, gluon condensate $\left\langle\alpha_sG_{\mu\nu}G^{\mu\nu}\right\rangle\,\approx\,0.007\,\text{GeV}^4$ and gluon mass $m_A=m/2=0.87\,\text{GeV}$.

We converted $F_{\mu\nu}F^{\mu\nu}$ into thermodynamic potential $\Phi$ coupled to the dielectric function $G(\phi)$. The $\Phi$ serves as a thermal correction representing quantum fluctuations around $V(\phi)$. Therefore, the $V_{eff}(\phi, T)$ leads to the restoration of the initially broken down symmetry in $V(\phi)$ at higher temperatures $T\geq T_{c\phi}$.  We classify the study into three parts: \textit{color confinement/deconfinement, dense nuclear matter made up of gluons and glueballs, and cosmological phase transitions}. In Sec.~\ref{CG}, we discussed the phenomenon of \textit{color confinement} through the derivation of a confining potential with a Cornell-like behavior. The $V_c(r)$ shows a linear rise from the asymptotically free region to the nonperturbative region where the particles are in a confined state. Despite the confining behavior observed at a relatively large inter-particle separation distance, hadronization sets in when the separation distance between the particles reaches a particular threshold. The hadronization phase is represented by vertical lines in Figs.~\ref{pba} and \ref{pd}. The \textit{deconfined matter phase} is consistent with modern-day cosmological data and the scenario of the inflationary universe. 

In Sec.~\ref{CG1} we study a \textit{dense nuclear matter consisting of gluons and glueballs}, a phenomenon that mimics the high baryon density regime of the early universe where matter was in a deconfined state. Here, we established that the presence of glueball fields $\chi$ in the glueball condensate reduces its quantity (see Fig.~\ref{pc}). Also, we find that in such a dense matter region, the associated \textit{effective glueball mass}, made up of both gluons and scalar glueballs, is unstable with negative square mass (see Fig.~\ref{pc2}), which inevitably makes the gluon condensate negative. The glueball potential in this region also lies in the asymptotically free medium with a negative gradient, Fig.~\ref{pc2a}. This phenomenon is consistent with the observations made in the QGP region, where the gluon condensate and isoscalar glueball masses change signs. We also studied pure gluon condensate with no glueballs and found that the condensate decreases with increasing temperature and vanishes at $T=T_{cg}$, (see Fig.~\ref{pc1}). {The variation of the scale factor with time as $T$ increases beyond $T_{c\phi}$ in Fig.~\ref{fig:at} shows that the inflationary expansion of the universe grows exponentially with increasing temperature, indicating strong glueball deconfinement.

In particular, the agreement between our model’s predictions and the Planck data, visible in Fig.~\ref{fig:example}, validates the core premise of our framework. By constraining the parameter $\gamma$ to obtain this fit, we show that the effective potential $V_{\text{eff}}(\phi, T)$, whose structure is rooted in QCD confinement dynamics, naturally supports a viable inflationary epoch across a physically reasonable parameter range. This robust match, achieved without significant fine-tuning, demonstrates that the gluon condensate, encoded through the parameters $\lambda$ and $m$, provides both the correct energy scale and the required dynamical behavior for inflation, with $\gamma$ serving as a meaningful bridge between the QCD and cosmological sectors. As a result, the scalar glueball field $\phi$ emerges as a credible inflaton candidate, offering an intrinsic connection between strong-interaction physics and the evolution of the early universe.

}

Additionally, to set the grounds for discussing the \textit{cosmological phase transitions}, we discussed the temperature correction to the scalar potential. In this section, we studied how the temperature correction modifies the glueball mass Fig.~\ref{pbb}, and leads to a phase transition at $T\geq T_{c\phi}$, Fig.~\ref{sp}. In Sec.~\ref{confinement} we unified the cosmological constants and \textit{color confining/deconfinement} phenomena. A $T=0$, we have \textit{color confinement} and at $T=T_{c\phi}$ there is the transition from \textit{confinement to deconfinement phase} with zero cosmological constants, and at $T>T_{c\phi}$ we have a \textit{deconfinement phase} with nonzero cosmological constant as well. Nonzero cosmological constant in the model framework at $T=0$ implies a universe full of fluctuating glueballs (see Fig.~\ref{pc1}).  

Furthermore, before the phase transition occurs at $T\,=\,T_{c\phi}$,  the required $T$ is in the order less than the electroweak phase transition energy i.e. $T_C\sim\, 100\, \text{GeV}$. The deconfinement temperature for gluons has been estimated to be around $T_{cg}\,\sim\,170\,\text{MeV}$ whilst that of fermions is estimated to be $T_{c\psi}\,\sim\,270\,\text{MeV}$ \cite{Fukushima, Cheng, Issifu}. We chose the value of the lightest glueball mass $m=1.73\,\text{GeV}$ determined from QCD phenomenology and QCD lattice calculations for the analysis \cite{Sexton, Morningstar, Loan, Chen}. The QCD vacuum for the Bag constant, which is known to be constant throughout space-time, is given as  $B_0=(300)^4\,\text{MeV}^4$ \cite{Brodsky}.

\subsection{Conclusion}
We have established a unified framework in which the dynamics of a scalar glueball field, $\phi$, rooted in nonperturbative QCD, simultaneously governs confinement and drives cosmological inflation. Using a color dielectric function \(G(\phi)\) coupled to an abelian gauge field, we connected low-energy QCD phenomenology with cosmological dynamics, revealing a coherent picture of confinement, phase transitions, deconfinement, and inflation.

Our main results are summarized below:

\begin{itemize}
    \item {\it Confinement and QCD Phenomenology.} From the static sector, we derived a Cornell-like confining potential $V_c(r)$, which shows asymptotic freedom at short distances and linear behavior at large separations. A hadronization threshold appears around $r \sim 1.85\,\text{GeV}^{-1}$.

    \item {\it Finite-Temperature Phase Transitions.} Integrating out the gauge field yields an effective potential $V_{eff}(\phi,T)$ with restored symmetry at high temperature $(T \ge T_{c\phi})$. The model predicts a clear thermal phase transition: glueballs are confined for $T<T_{c\phi}$, massless at $T = T_{c\phi}$, and deconfined for $T > T_{c\phi}$, consistent with lattice QCD behavior.

    \item {\it Dense Glueball–Gluon Matter.} Within the relativistic mean field approximation, dense glueball–gluon matter exhibits a negative effective glueball mass squared, signaling deconfinement. The gluon condensate decreases with temperature and vanishes at a critical value $T_{cg}$, with the glueball mean field further suppressing it.

    \item {\it Cosmological Implications.} The glueball field also acts as the inflaton, linking the gluon condensate to the vacuum energy and cosmological constant $\Lambda$. For $T > T_{c\phi}$, a positive $\Lambda$ drives inflation; at $T = T_{c\phi}$, $\Lambda$ vanishes; and at $T=0$, the universe settles into a confined phase with a small positive cosmological constant. Domain-wall issues are naturally resolved through high-temperature symmetry restoration.

    \item {{\it Consistency with Cosmological Data:} As shown in Fig.~\ref{fig:example}, the predicted values of $n_s$ and $r_s$ lie within the Planck bounds, yielding physically reasonable constraints on $\gamma$ and confirming that the glueball-based QCD-inspired potential can naturally generate a viable inflationary epoch. This agreement demonstrates that the model effectively links nonperturbative strong-interaction physics with precision cosmological observations.}

\end{itemize}

Overall, we developed a unified framework in which the nonperturbative dynamics of QCD, encoded through a glueball field coupled to an effective gauge sector, simultaneously generate color confinement and a viable inflationary phase. By integrating out the gauge field and incorporating thermal corrections into the effective potential $V_{eff}(\phi,T)$, we showed that the early universe undergoes a first-order phase transition at $T = T_{c\phi}$, whose supercooled vacuum structure drives the onset of inflation. The resulting slow-roll dynamics yield predictions for $n_s$, $r_s$, and $\alpha_s$ that align with Planck data for physically reasonable values of the parameter $\gamma$ and $N \approx 50\text{--}60$, with the characteristic relation $r_s = 4(1-n_s)-72\gamma$ providing a distinct observational signature. Normalizing the scalar amplitude further constrains $\sigma^2$, linking cosmological observables to QCD-scale physics. These results firmly position the scalar glueball field as a compelling inflaton candidate and provide a testable bridge between strong-interaction physics and early-universe cosmology, inviting future work on gravitational waves, phase transition signatures, and further cross-disciplinary constraints.

\acknowledgments
This work is partly supported by Conselho Nacional de Desenvolvimento Cient\'ifico e Tecnol\'ogico (CNPq) project No.: 168546/2021-3, Brazil. A. I. acknowledges financial support from the São Paulo State Research Foundation (FAPESP), Grant No. 2023/09545-1 and F. A. B. acknowledges support from CNPq,  Grant No. 309092/2022-1. This work is part of the project INCT-FNA (Proc. No. 464898/2014-5) and is also supported by the National Council for Scientific and Technological Development (CNPq) under Grants No.  306834/2022-7 (T.F.). T. F. also thanks the financial support from  Improvement of Higher Education Personnel CAPES (Finance Code 001) and FAPESP Thematic Grants (2023/13749-1 and 2024/17816-8)
\\
\\
Data Availability Statement: The datasets used and/or analyzed during the current study are available from the corresponding author upon reasonable request. 

\bibliography{reference}

\end{document}